\documentclass[pdflatex,sn-nature]{sn-jnl}
\usepackage{graphicx}
\usepackage{amsmath,amssymb,amsfonts}
\usepackage{physics}
\usepackage{bm}
\usepackage{braket}
\usepackage{xcolor}
\usepackage{booktabs}
\usepackage{needspace}
\usepackage{tikz}
\graphicspath{{figures/}}

\newcommand{\f}{\frac}
\newcommand{\ep}{\epsilon}

\newcommand{\ta}{\theta}
\newcommand{\p}{\partial}
\newcommand{\al}[1]{\langle #1 \rangle}

\providecommand{\dg}{\dagger}

\providecommand{\alg}[1]{\begin{align}#1\end{align}}
\hypersetup{pdftitle={Thermal Hall effect induced by phonon skew scattering via orbital magnetization},pdfauthor={Taekoo Oh},pdfsubject={Phonon thermal Hall effect}}
\begin{document}
\allowdisplaybreaks[0]
\title{Thermal Hall effect induced by phonon skew scattering via orbital magnetization}
\author*[1]{\fnm{Taekoo} \sur{Oh}}\email{taekoo.oh@ssu.ac.kr}
\affil*[1]{\orgdiv{Department of Physics}, \orgname{Soongsil University}, \orgaddress{\city{Seoul}, \postcode{06978}, \country{Korea}}}
\abstract{
The thermal Hall effect in insulators without spin order motivates spin-independent mechanisms of phonon skew scattering.
Here, I propose a new mechanism by which time-reversal-broken electronic orbital motion generates an orbital-magnetization--phonon (OMP) Raman coupling to phonon angular momentum.
A spinless Haldane model gives the spatial structure of the coupling and its distinct relation to orbital circulation across a topological phase transition.
Random modulations of the local OMP coupling provide dilute scattering centers, and the effective Raman coupling energy can be determined from the impurity-induced phonon full width at half maximum (FWHM).
An acoustic phonon Boltzmann calculation shows that a representative FWHM of $0.0119$ meV at a concentration of $0.01$ per primitive cell gives a coupling energy near $10$ meV and a Hall angle of $8.09\times10^{-4}$ at $100$ K.
This work advances the understanding of electron–phonon coupling and provides a pathway to tailor thermal properties in various materials.
}
\maketitle
\section{Introduction}
In insulators, thermal transport is a powerful probe of elementary excitations, as the charge gap blocks electronic conduction.
A notable thermal Hall effect (THE) has been observed experimentally in
magnetic and Mott insulators, Kitaev candidates, and cuprates~\cite{strohm2005phenomenological,inyushkin2007phonon,onose2010observation,mori2014origin,hirschberger2015large,hirschberger2015thermal,ideue2017giant,zhang2021anomalous,akazawa2022topological,kim2024thermal,kasahara2018majorana,hentrich2018unusual,kasahara2018unusual,hentrich2019large,lefranccois2022evidence,bruin2022robustness,chen2024planar,grissonnanche2019giant,boulanger2020thermal}.
 Spin and phonon mechanisms can both contribute.
While early works emphasized a spin-based origin~\cite{katsura2010theory,matsumoto2011rotational,owerre2016first,owerre2017topological,zhang2019thermal,zhang2021topological,zhang2024thermal}, phonon contributions are now considered equally important~\cite{sheng2006theory,kagan2008anomalous,wang2009phonon,zhang2010topological,agarwalla2011phonon,qin2012berry,saito2019berry,oh2025phonon,oh2025spin}.

For phonons, a Raman interaction $\mathbf{M}\!\cdot\!\mathbf{L}$ with a time-reversal-odd magnetization can produce an intrinsic Berry-curvature response or extrinsic skew scattering by scalar spin chirality, defects, or collective excitations~\cite{sheng2006theory,kagan2008anomalous,wang2009phonon,zhang2010topological,agarwalla2011phonon,qin2012berry,saito2019berry,oh2025phonon,mangeolle2022phonon,sun2022large}.
Here, $\mathbf{M}$ is the magnetization and $\mathbf{L}=\mathbf{u}\times\mathbf{p}$ is the phonon angular momentum, with ionic displacement $\mathbf u$ from equilibrium and conjugate momentum $\mathbf p$.
In particular, the extrinsic mechanism through scalar spin chirality successfully explains the temperature dependence of the THE in Mott insulators and Kitaev candidates.

Remarkably, substantial THE signals were recently reported also in non-magnetic insulators and semiconductors, including SrTiO$_3$, Bi$_{2-x}$Sb$_x$Te$_{3-y}$Se$_y$, Y$_2$Ti$_2$O$_7$, black phosphorus, MgO, MgAl$_2$O$_4$, SiO$_2$, Si, and Ge~\cite{li2020phonon,sim2021sizable,sharma2024phonon,sharma2024phonon2,li2023phonon,jin2024discovery}.
Typical thermal Hall angles are $|\theta_H|=|\kappa_{yx}/\kappa_{xx}|\sim10^{-3}$, where $\kappa_{xx}$ and $\kappa_{yx}$ are the longitudinal and transverse thermal conductivities for a temperature gradient along $x$. The measured conductivities scale as $|\kappa_{yx}|\propto \kappa_{xx}$ or $\kappa_{xx}^2$.
This motivates examining a spin-independent, time-reversal-odd source of extrinsic phonon scattering beyond a spatially uniform intrinsic response.

Orbital magnetization (OM)~\cite{thonhauser2005orbital,xiao2005berry,ceresoli2006orbital,shi2007quantum,xiao2010berry,thonhauser2011theory,aryasetiawan2019modern}, generated by electronic circulation, provides such a source but has received less attention in relation to phonons.
Therefore, in this work, I propose a new mechanism applicable to various insulators and semiconductors: phonon skew scattering mediated by OM.
Here, I use the spinless Haldane model with internal time-reversal breaking to demonstrate the mechanism~\cite{haldane1988model}.
With the Born--Oppenheimer approximation, I first derive the corresponding orbital-magnetization--phonon (OMP) Raman coupling and the molecular Berry curvature $b_i^z$ that generates it.
I reproduce the established Haldane phase diagram in a periodic system, and use L\"owdin orbitals of a finite flake to evaluate the local circulation (LC), itinerant circulation (IC), and total OM~\cite{thonhauser2005orbital,thonhauser2006insulator}.
The clean finite-flake calculation finds that $b_i^z$ depends strongly on position, differing between bulk and edge sites and between the $A$ and $B$ sublattices.
For transport, I model spatial modulations of the local OMP coupling as randomly positioned dilute scattering centers, with their effective strength estimated from the phonon linewidth.
The resulting Hall response is treated as extrinsic skew scattering, not as the intrinsic phonon thermal Hall response of a spatially uniform OMP coupling.

The site-averaged molecular Berry curvature $b^z = \f{1}{N_{\rm site}}\sum_i b_i^z$ shows different behavior for trivial and Chern insulating phases. Here, $i$ labels a lattice site, $N_{\rm site}$ is the number of sites, and $z$ is the out-of-plane direction.
In a trivial insulator, $b^z$ associated with the OMP coupling closely follows the LC.
In a Chern insulator, $b^z$ deviates from the LC and develops a plateau-like feature correlated with the reconstruction of bulk and edge states.
Using a scattering-center concentration of about $1~\%$ per primitive cell and an impurity-induced phonon full width at half maximum (FWHM) guided by acoustic-phonon experiments, I estimate an effective Raman coupling energy of about 10~meV.
The resulting local scattering model estimates a Hall angle near $10^{-3}$ at 100~K.

Even in a spinless system, time-reversal-breaking electronic orbital motion can generate an OMP coupling whose spatial variations drive phonon skew scattering.
This motivates a search for the same mechanism in materials with OM, including Mott insulators~\cite{lovesey2002orbital}, chiral $p$-wave superconductors~\cite{braude2006orbital,holmvall2023enhanced}, itinerant magnets~\cite{meyer1961experimental,eriksson1990orbital,hiess2001spin,solovyev2005orbital,ceresoli2010first,ovesen2024orbital}, and $p$-type semiconductors~\cite{cheng2009orbital,sliwa2014orbital,ovesen2024orbital}.
Also, an applied magnetic field can be described by Peierls phases in a tight-binding model, giving rise to OM.
Hence, OM can be induced by both internal orbital phases and an applied magnetic field, further supporting the broad applicability of the mechanism.

\section{Results}
\subsection{Introduction to the Raman interaction}
Because electronic energy scales are generally much higher than nuclear (ionic) energy scales, one first determines the electronic eigenstates with fixed nuclear positions.
This is the Born--Oppenheimer (BO) approximation.
For fixed ionic coordinates $\{\mathbf{R}_i\}$, let $|\Psi_{\mathrm{el}}(\{\mathbf{R}_i\})\rangle$ be the electronic many-body ground state. Integrating out the electronic degrees of freedom gives the effective nuclear Hamiltonian
\begin{align}
H_{\mathrm{nu}}^{\mathrm{eff}}
&=E_{\mathrm{el}}(\{\mathbf{R}_i\})+
\sum_i\frac{(\mathbf{P}_i-\hbar\boldsymbol{\mathcal A}_i)^2}{2M_i},
\nonumber\\
\boldsymbol{\mathcal A}_i
&=i\langle\Psi_{\mathrm{el}}|
\boldsymbol{\nabla}_{R_i}|\Psi_{\mathrm{el}}\rangle .
\label{eq:BO}
\end{align}
Here, $E_{\rm el}$ is the electronic ground-state energy at fixed ionic coordinates, $\mathbf P_i$ is the momentum conjugate to $\mathbf R_i$, and $M_i$ is the mass of ion $i$.
The molecular Berry connection $\boldsymbol{\mathcal A}_i$ arises from the dependence of the electronic wavefunction on ionic displacements.
Broken time-reversal symmetry permits a finite molecular Berry curvature {\mbox{$b_i^z=\p_{u_i^x}\mathcal A_i^y-\p_{u_i^y}\mathcal A_i^x$}} in {\it real space}, which produces the local Raman coupling
\begin{align}
H_R&=-\sum_i\ep_i^{\mathrm{BO}}\frac{L_i^z}{\hbar},
\qquad
\ep_i^{\mathrm{BO}}=\frac{\hbar^2b_i^z}{2M_i},
\nonumber\\
L_i^z&=(\mathbf{u}_i\times\mathbf{p}_i)_z.
\label{eq:Raman}\end{align}
Here, $\mathbf u_i$ is the displacement of ion $i$ from equilibrium, $\mathbf p_i=\mathbf P_i$ is its conjugate momentum, and $\epsilon_i^{\rm BO}$ is the local BO Raman coupling energy.
This coupling plays a key role in the phonon THE through both intrinsic and extrinsic mechanisms.
$b_i^z$ can arise from various sources, such as the overlap of atomic orbitals~\cite{saito2019berry}, ionic coupling to an external magnetic field~\cite{agarwalla2011phonon}, the coupling to magnetic order via spin-orbit interaction or scalar spin chirality induced by spin fluctuations~\cite{oh2025phonon,oh2025spin}. Here, I examine the orbital electronic source in a spinless model.

\subsection{OMP coupling exemplified in the Haldane model}
\begin{figure}[t]
     \centering
       \includegraphics[width=.94\linewidth]{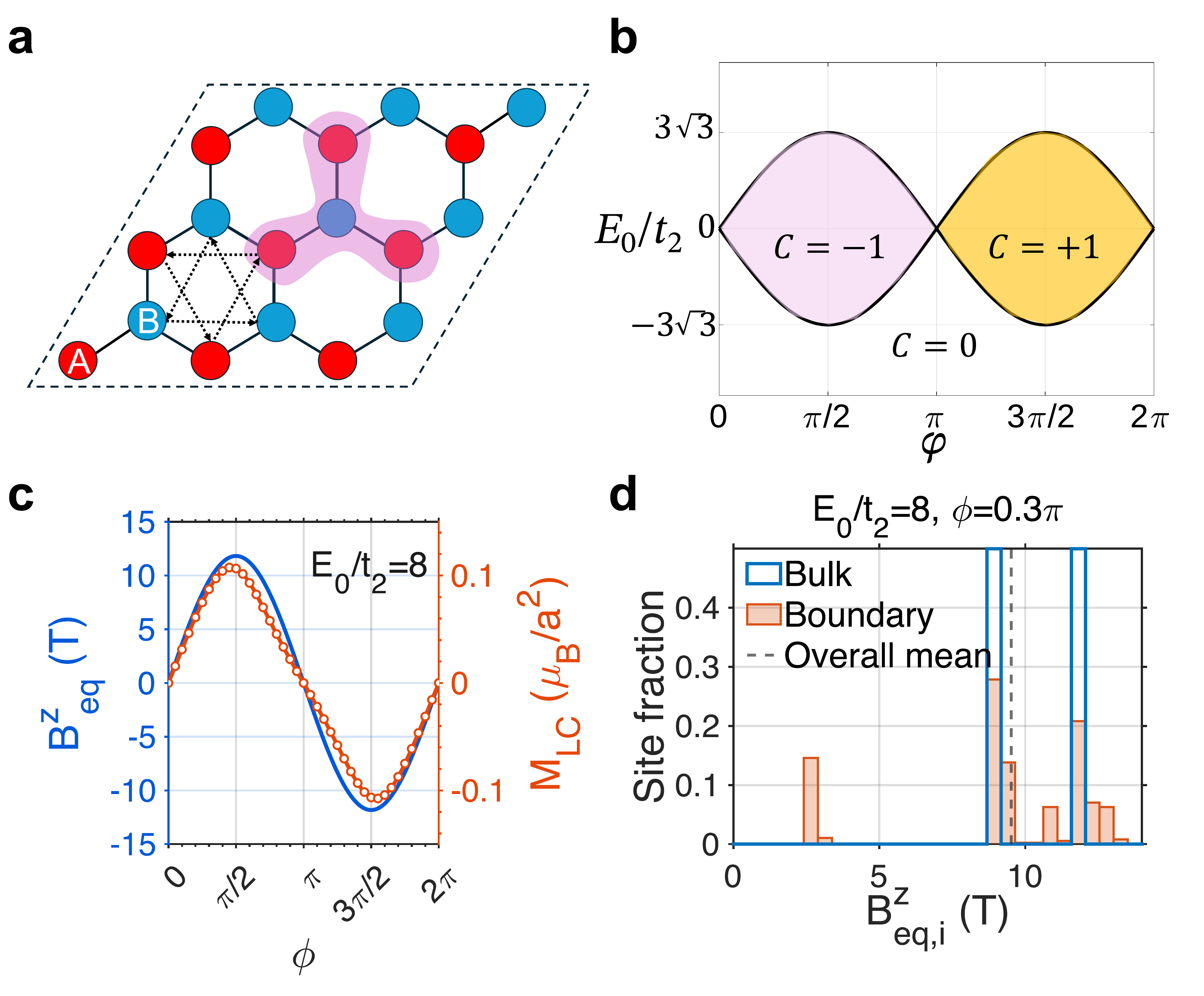}
     \caption{\textbf{Orbital magnetization and equivalent field in the Haldane model.}
     (a) Haldane lattice and a representative L\"owdin-localized orbital.
     (b) Phase diagram for $|t_2/t_1|<1/3$.
     (c) Spatially averaged finite-flake equivalent field $B_{\mathrm{eq}}^z=(\hbar/e)b^z$ and the central-orbital LC $M_{\mathrm{LC}}$ versus $\varphi$ at $E_0/t_2=8$.
     (d) Normalized bulk and boundary distributions of $B_{\mathrm{eq},i}^z$ at $E_0/t_2=8$ and $\varphi=0.3\pi$. The boundary distribution is filled to the baseline and the gray dashed line is the overall mean. SI Fig.~S1 shows the corresponding real-space map.
     }
       \label{fig:haldane}
 \end{figure}

I use the spinless Haldane model and consider a finite flake [Fig.~\ref{fig:haldane}(a)].
$E_0>0$ is half the onsite-energy difference between the sublattices: the onsite energies are $+E_0$ on $A$ and $-E_0$ on $B$.
The onsite energy on sublattice $B$ is chosen to be lower than that on sublattice $A$ without loss of generality.
The nearest-neighbor (NN) hopping $t_1$ carries no additional phase, whereas the next-nearest-neighbor (NNN) hopping $t_2$ carries a phase $\varphi$, which represents an internal time-reversal-breaking flux.
The net flux per unit cell vanishes.
For an infinite periodic system, I reproduce the phase diagram~\cite{thonhauser2005orbital} for $|t_2/t_1|<1/3$.
For the finite-flake results, I use $t_1=1$~eV, $t_2=t_1/4$, and a triangular Bravais-lattice constant $a=5$~\AA{}.
The flake contains $16\times16$ primitive cells, or $N_{\mathrm{site}}=512$ sites, and is half-filled.
The model is described in Methods.

To evaluate OM, I represent the occupied electronic subspace by orthonormal L\"owdin orbitals $|\phi_n\rangle$, obtained by projecting trial orbitals on the lower-energy sublattice $B$ into the occupied subspace and symmetrically orthogonalizing them.
The orbital moment of each occupied orbital and the total OM per unit area are
\begin{align}
 m_{n,z}&=-\frac{e}{2}\langle\phi_n|(\mathbf r\times\mathbf v)_z|\phi_n\rangle,
 \nonumber\\
 M^{\mathrm{tot}}&=\frac{1}{A}\sum_{n\in\mathrm{occ}}m_{n,z}.
 \label{eq:OM-results}
\end{align}
Here $e>0$ is the elementary charge, $\mathbf r$ is the position operator, and $\mathbf v=-(i/\hbar)[\mathbf r,H]$ is the velocity operator for the electronic Hamiltonian $H$.
The flake area is $A=16^2A_0$, with primitive-cell area $A_0=\sqrt3a^2/2$.
All OM quantities below refer to the out-of-plane component per unit area, plotted in units of $\mu_B/a^2$, where $\mu_B$ is the Bohr magneton.

\begin{figure}[t]
     \centering
     \includegraphics[width=\linewidth]{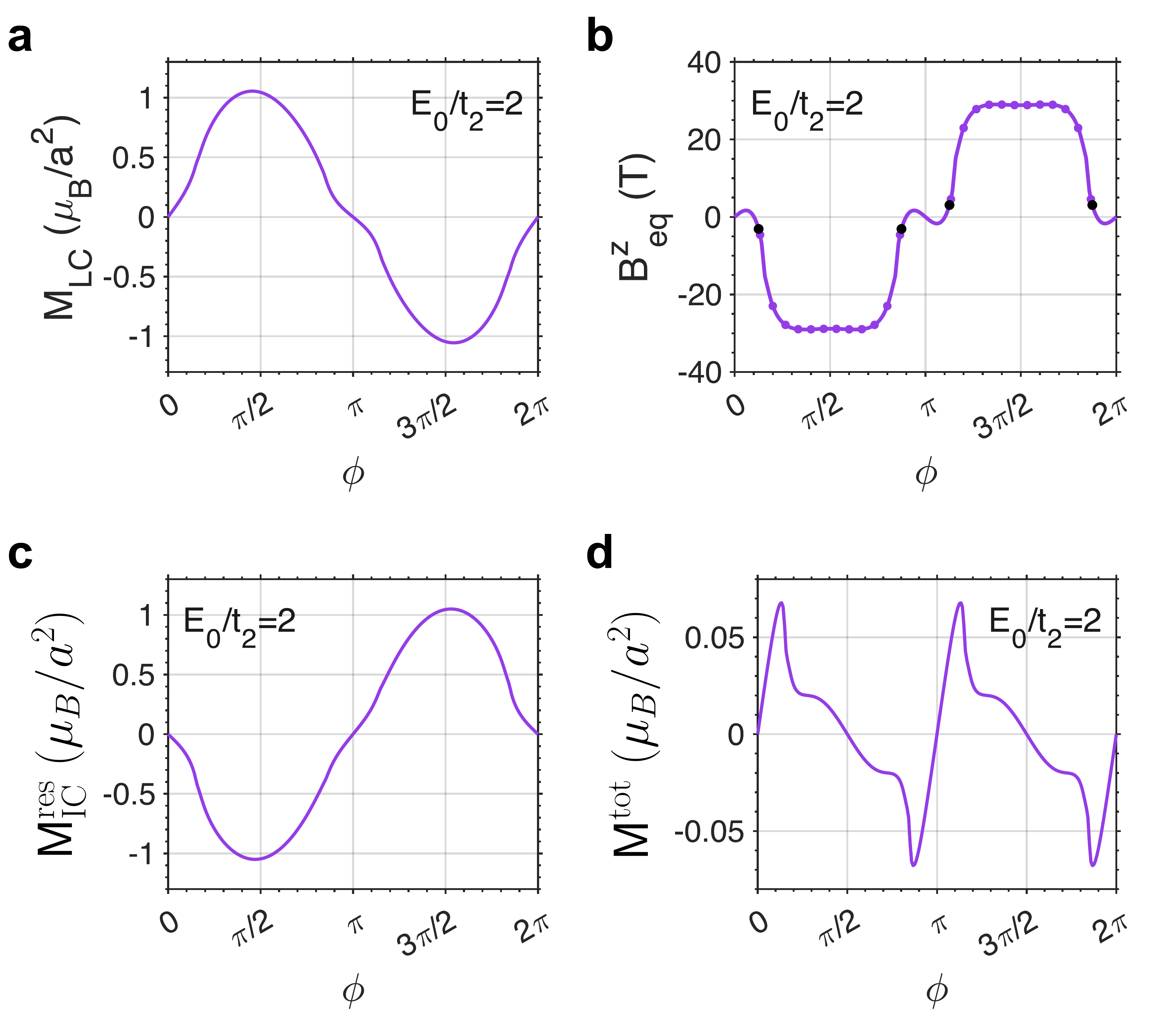}
     \caption{ \textbf{Orbital-magnetization and equivalent-field curves at $E_0/t_2=2$.}
    (a),(b) Finite central-orbital $M_{\mathrm{LC}}$ and spatially averaged finite-flake $B_{\mathrm{eq}}^z$.
    (c),(d) The operational finite residual $M_{\mathrm{IC}}^{\mathrm{res}}\equiv M^{\mathrm{tot}}-M_{\mathrm{LC}}$ and finite total OM $M^{\mathrm{tot}}$.
    The corresponding curves for $E_0/t_2=1,3,4$, together with the $E_0/t_2=8$ OM comparison in (c),(d), are shown in SI Fig.~S3.
    All curves use an exactly half-filled $16\times16$ open flake with all three next-nearest-neighbor bond directions included. Black dots in (b) are evaluated directly at the analytic phase boundaries; colored dots lie in the Chern-insulating region. }
     \label{fig:haldane-curves}
 \end{figure}

Local circulation (LC) describes the self-rotation of an orbital about its center $\bar{\mathbf r}_n=\langle\phi_n|\mathbf r|\phi_n\rangle$, whereas itinerant circulation (IC) describes circulation of the orbital centers.
The local moment due to self-rotation is
\begin{equation}
 m_{n,z}^{\mathrm{LC}}=-\frac{e}{2}\langle\phi_n|[(\mathbf r-\bar{\mathbf r}_n)\times\mathbf v]_z|\phi_n\rangle.
 \label{eq:LC-results}
\end{equation}
The IC is defined by $m_{n,z}-m_{n,z}^{\mathrm{LC}}$.
For Figs.~\ref{fig:haldane} and~\ref{fig:haldane-curves}, I estimate the bulk LC from a L\"owdin orbital $c$ near the flake center. The LC estimate and the residual IC (superscript $\mathrm{res}$) are
\begin{align}
 M_{\mathrm{LC}}&\equiv \frac{m_{c,z}^{\mathrm{LC}}}{A_0},
 \nonumber\\
 M_{\mathrm{IC}}^{\mathrm{res}}&\equiv M^{\mathrm{tot}}-M_{\mathrm{LC}}.
 \label{eq:OM-figure-definitions}
\end{align}
As discussed in Section~S2 of the Supplementary Information (SI), this residual can differ in a finite flake from the sum of the itinerant orbital moments divided by $A$.
Computational details are provided in Methods.

For comparison with OM, I express the molecular Berry curvature as the equivalent field $B_{\mathrm{eq},i}^z=(\hbar/e)b_i^z$ and its site average $B_{\mathrm{eq}}^z=N_{\mathrm{site}}^{-1}\sum_iB_{\mathrm{eq},i}^z$.
Here $i$ labels a lattice site, while $n$ above labels an occupied orbital.
The equivalent field expresses the molecular Berry curvature in tesla.

Figures~\ref{fig:haldane}(c,d) and~\ref{fig:haldane-curves} summarize the key results.
First, the trivial and Chern regimes show different relations between OM and molecular Berry curvature.
In the trivial regime, $B_{\mathrm{eq}}^z=\f{1}{N_{\rm site}}\sum_i B_{\mathrm{eq},i}^z$ closely follows $M_{\mathrm{LC}}$ [Fig.~\ref{fig:haldane}(c)].
However, in the Chern regime, $B_{\mathrm{eq}}^z$ changes sign and develops a plateau-like dependence on $\varphi$ [Fig.~\ref{fig:haldane-curves}].
Neither $M_{\mathrm{IC}}^{\mathrm{res}}$ nor the finite total $M^{\mathrm{tot}}$ shows the same profile.

The sign reversal and plateau-like profile are associated with the reconstruction of the finite-flake electronic states across the topological transition, although the plateau height is not quantized.
The site maps in SI Fig.~S2 show that the redistribution of molecular Berry curvature begins in the interior and subsequently reaches the boundary.
Thus, the deviation from the LC trend is correlated with the bulk--edge reconstruction accompanying the chiral edge mode~\cite{hasan2010colloquium}.
A useful qualitative picture is that localized orbital circulation cancels substantially in the interior while an uncompensated boundary circulation remains.

Next, I distinguish the clean local OMP profile from the extrinsic scattering ensemble.
Figure~\ref{fig:haldane}(d) shows that the local molecular Berry curvature is spatially nonuniform even in the clean flake, with different bulk and boundary distributions.
The standard deviation is nearly $30~\%$ of the mean even in the clean flake.
However, the clean profile does not itself define a disorder ensemble. The onsite-disorder calculation in SI Section~S4.1 examines how local electronic circulation responds to a perturbed environment.

In the transport calculation below, I model the disorder as a separate ensemble of randomly positioned dilute OMP scattering centers and average the scattering rates over their positions.
To specify the local coupling profile, let $\bar b_\mu^z$ be the mean molecular Berry curvature on sublattice $\mu=A,B$ within the central $8\times8$ cells of the $16\times16$ flake.
The dimensionless sublattice weights are defined as $s_\mu=2\bar b_\mu^z/(\bar b_A^z+\bar b_B^z)$, so that $(s_A+s_B)/2=1$.
For example, $E_0/t_2=8$ and $\varphi=0.3\pi$ give $s_A=0.86$ and $s_B=1.14$.
These weights set the relative coupling energies on the two sublattices in the transport calculation.

\subsection{Estimating the coupling energy from the phonon FWHM}

The Haldane calculation identifies the orbital mechanism and provides representative sublattice weights.
However, a single electronic model cannot describe the full variety of defects and local environments in real materials.
Therefore, I use an effective scattering model based on phonon parameters from real materials and the impurity-induced linewidth.
Here, I adopt a local coupling profile and a scattering-center concentration, and infer the effective coupling magnitude $|\epsilon_\star|$ from a representative phonon FWHM.

I denote the single-center vertex by $U_{fi}=\epsilon_\star\gamma_{fi}$, where $\epsilon_\star$ is the effective Raman coupling energy, $\gamma_{fi}$ is the dimensionless weight connecting $i$ and $f$ states, and $i$ and $f$ label the initial and final acoustic phonon states in momentum space.
Each state label includes momentum and acoustic branch, and $E_i$ and $E_f$ are the initial and final phonon energies. The factor $\gamma_{fi}$ contains the phonon polarizations, energy ratios, and the sublattice profile (see Methods).
For dilute independent OMP scattering centers, the impurity-induced phonon FWHM is
\begin{align}
 \Delta E_{{\rm FWHM},i}^{\rm imp}\equiv\Gamma_i^{\rm imp}=2\pi c_{\rm imp}\epsilon_\star^2\mathcal K_i
 \label{eq:linewidthcalibrationmain}
\end{align}
Here, $\Gamma_i^{\rm imp}$ is the impurity-induced energy linewidth, equal to $\hbar$ times the second-order scattering rate. The linewidth kernel is $\mathcal K_i \equiv \sum_{f\ne i}|\gamma_{fi}|^2\delta(E_f-E_i)/N_k$, where $N_k$ is the number of momentum points in the grid. The scattering-center concentration per primitive cell is $c_{\rm imp}=N_c/N_{\rm cell}$, with $N_c$ scattering centers in $N_{\rm cell}$ primitive cells. Computational details are provided in Methods.

To estimate the coupling energy and calculate transport, I use longitudinal and transverse spring constants $K_{\parallel}=21.998$ and $K_{\perp}=5.010~\mathrm{eV}\,\text{\AA}^{-2}$, and sublattice masses $M_A=10.81$ and $M_B=14.00$~u, as used for $h$-BN~\cite{michel2009theory}. Figure~\ref{fig:transport}(a) shows the phonon spectrum.
In the $50\times50$ $k$-grid calculation,
the median of $\mathcal K_i$ over acoustic states in the 8--12-meV energy window is $1.89\times10^{-3}$ meV$^{-1}$.
Guided by experimental scales~\cite{kulda2001thermal,chou2019long}, I choose a representative impurity-induced phonon FWHM of $0.0119$ meV. At $c_{\rm imp}=0.01$ per primitive cell, this gives $|\epsilon_\star|=10.0$ meV.
This is an illustrative scale rather than a measurement from a specific sample.
For a gapped insulator with $k_BT$ well below the electronic gap, thermal occupations change the OM only weakly~\cite{shi2007quantum}, so a temperature-independent OMP scale is reasonable.

\subsection{Phonon thermal Hall effect via skew scattering by OM}

Using the coupling energy estimated above, I calculate the phonon thermal Hall effect generated by dilute random OMP scattering centers.
These centers are assumed to provide the only scattering channel.
The second-order symmetric and third-order antisymmetric rates enter the Boltzmann equation. The two acoustic branches are used in the initial, final, and virtual sums. I report the concentration-normalized conductivities $\hat\kappa_{\alpha x}=c_{\rm imp}\kappa_{\alpha x}$ for $\alpha=x,y$. The superscript $\mathrm{odd}$ denotes the component odd under reversal of the OMP coupling. The Hall angle is independent of this concentration at fixed $\epsilon_\star$ within the OMP-only dilute model.

\begin{figure}[t]
    \centering
       \includegraphics[width=\linewidth]{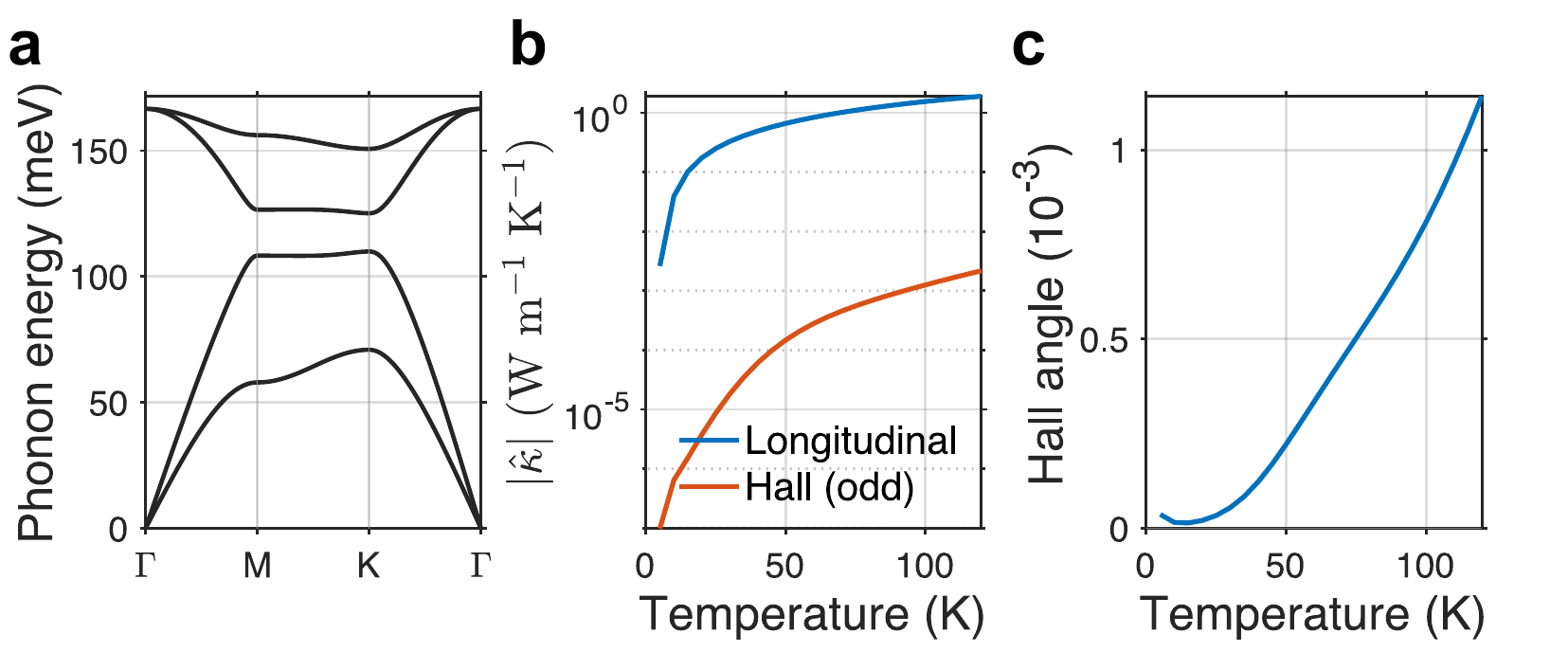}
     \caption{\textbf{Acoustic-phonon transport from OMP skew scattering.}
     (a) Phonon spectrum of the honeycomb spring model; only the two acoustic branches enter the transport calculation.
     (b) Concentration-normalized dilute coefficients $\hat\kappa_{xx}\equiv\lim_{c_{\mathrm{imp}}\to0}c_{\mathrm{imp}}\kappa_{xx}$ and $|\hat\kappa_{yx}^{\mathrm{odd}}|$. (c) $\ta_H=\kappa_{yx}^{\mathrm{odd}}/\kappa_{xx}$; this curve is unchanged by the concentration normalization.
     The reference coupling energy is $\epsilon_\star=10$~meV.}
       \label{fig:transport}
 \end{figure}

For $\ep_\star=10$~meV, the relation $\epsilon_\mu^{\rm tr}=\epsilon_\star s_\mu$ gives the sublattice transport coupling energies: $\ep_A^{\rm tr}=8.61$ and $\ep_B^{\rm tr}=11.39$~meV. Here, $N$ denotes the number of momentum points along each grid direction: $N_k=N^2$.
At $100$~K the $N=50$ calculation gives
\begin{align}
\hat\kappa_{xx}
&=1.55~\mathrm{W\,m^{-1}K^{-1}},
\nonumber\\
\hat\kappa_{yx}^{\mathrm{odd}}
&=1.25\times10^{-3}~\mathrm{W\,m^{-1}K^{-1}},
\nonumber\\
\ta_H&=8.09\times10^{-4}.
\label{eq:transport100K}
\end{align}
At fixed $\epsilon_\star$, the absolute conductivities require an independently specified scattering-center concentration, whereas their ratio does not within the present OMP-only collision model.
Checks with $N=20,30,40,50$ and numerical energy resolution $\eta=2$--$4$~meV give the same Hall-angle scale (see SI Section~S10).
The concentration-normalized symmetric and antisymmetric rates scale as $\hat W^{(2)}\propto\epsilon_\star^2$ and $\hat W^{(3a)}\propto\epsilon_\star^3$, respectively, giving $\theta_H\propto\epsilon_\star$.

Figures~\ref{fig:transport}(b,c) show $\hat \kappa_{xx}$, $|\hat \kappa_{yx}^{\rm odd}|$, and $\theta_H$ as functions of temperature.
The calculated angle increases from $2.15\times10^{-5}$ at 20~K to $8.09\times10^{-4}$ at 100~K.
These calculations use a fixed $\epsilon_\star$, but a sufficiently strong thermal reduction of $\epsilon_\star$ could produce a maximum in $\ta_H$~\cite{shi2007quantum,tserkovnyak2015spin}.

\section{Discussion}

The central result is an extrinsic phonon thermal Hall mechanism rooted in local orbital electronic circulation.
The Haldane calculation shows that the OMP coupling is spatially nonuniform and exhibits distinct behavior across the topological phase transition.
The phase diagram, trends in LC, IC, and total OM, and plateau-like molecular Berry curvature $b^z$ establish the electronic origin and spatial structure of the coupling and its correlation with the chiral bulk--edge reconstruction.
With the dilute OMP scattering centers, the effective coupling energy and associated thermal Hall angle are estimated.

Candidate materials should exhibit local time-reversal breaking and a spatially modulated OMP coupling.
$A$Fe$_2$(PO$_4$)$_2$ ($A={}$Ba, Cs, K, La) is a candidate material proposed to realize the Haldane model~\cite{kim2017realizing}.
SrTiO$_3$~\mbox{\cite{urazhdin2025atomic}}, Bi$_{2-x}$Sb$_x$Te$_{3-y}$Se$_y$~\mbox{\cite{kim2016unexpected,cullen2025giant}}, and Y$_2$Ti$_2$O$_7$, in which OM can be enhanced by spin--orbit coupling, are other possible candidates.
Orbital Hall studies~\mbox{\cite{gobel2024orbital}} also motivate examining Si~\mbox{\cite{matsumoto2025observation}}, Ge~\mbox{\cite{santos2024negative}}, and black phosphorus~\mbox{\cite{cysne2023ultrathin}}.
Since OM introduces a new scattering source for phonons, even some metals that violate the Wiedemann--Franz law, such as ferromagnetic Weyl semimetal Co$_2$MnAl~\mbox{\cite{robinson2021large}}, may also exhibit a sizable contribution to THE from phonons.
A spatially resolved phonon angular momentum Hall response may provide an additional experimental test of this mechanism~\mbox{\cite{park2020phonon}}.

The proposed extrinsic mechanism can be contrasted with the intrinsic response.
Skew scattering depends on fluctuations of the local OMP coupling, while the intrinsic response depends on the spatially periodic mean of the coupling.
The relative magnitudes of the intrinsic and extrinsic contributions depend on the mean coupling and disorder statistics.
Furthermore, at fixed temperature and coupling, varying the dilute-center concentration gives $|\kappa_{yx}|\propto\kappa_{xx}$ when OMP scattering dominates, whereas varying dominant background relaxation at fixed OMP scattering gives $|\kappa_{yx}|\propto\kappa_{xx}^2$~\cite{nagaosa2010anomalous,oh2025phonon}. 
On the other hand, $|\kappa_{yx}|$ is independent of $\kappa_{xx}$ for the intrinsic response.

Saito et al.~\cite{saito2019berry} described an intrinsic phonon thermal Hall effect arising from the atomic orbital response to an external magnetic field.
Here, I identify a new extrinsic mechanism in which electronic orbital circulation generates a local OMP coupling and its spatial modulations produce phonon skew scattering.
The spinless Haldane model shows that this mechanism can operate with internal time-reversal breaking, without the spin degrees of freedom required by scalar-spin-chirality mechanisms~\cite{oh2025phonon}.

Because OM occurs in Mott insulators, itinerant magnets, orbital-active semiconductors, and other time-reversal-broken electronic systems~\cite{lovesey2002orbital,meyer1961experimental,eriksson1990orbital,hiess2001spin,solovyev2005orbital,ceresoli2010first,cheng2009orbital,sliwa2014orbital}, the mechanism is not restricted to a particular magnetic order. Its requirements are local time-reversal breaking, an OMP coupling, and spatial modulation capable of producing skew scattering.
Hence, this work will help engineer the thermal properties of a wide range of materials by revealing the role of electron-phonon couplings in thermal transport.

\pagebreak

\section{Methods}
\subsection{Electronic model and molecular Berry curvature}

I use the spinless Haldane model~\cite{haldane1988model},
\alg{
H = [\sum_{\al{ij}} t_{1} c_i^\dg c_j + \sum_{\al{\al{ij}}} t_{2}e^{i\varphi_{ij}} c_i^\dg c_j] +\text{H.c.} + \sum_i \xi_i c_i^\dg c_i. \label{eq:Haldane}
}
Here $c_i$ annihilates a spinless electron at site $i$, and $\al{ij}$ and $\al{\al{ij}}$ denote nearest- and next-nearest-neighbor bonds, respectively. For counterclockwise hopping, $\varphi_{ij}=\varphi$, and the onsite energies are $\xi_A=-\xi_B=E_0>0$.
I set $t_1=1$ eV, $t_2=t_1/4$, and the triangular Bravais-lattice constant $a=5$ \AA{}.
The phase diagram uses the periodic model; local quantities use a $16\times16$-cell open flake with the lowest 256 of 512 electronic states occupied.
Only the three incident nearest-neighbor hoppings respond to an ionic displacement:
\begin{equation}
 \delta t_{ij}=g\hat{\mathbf d}_{ij}\cdot
 (\mathbf u_j-\mathbf u_i),\qquad
 g=0.1~\mathrm{eV}\,\text{\AA}^{-1}.
 \label{eq:dtdu}
\end{equation}
The unit vector $\hat{\mathbf d}_{ij}$ follows the original bond.
The derivative $g$ is an illustrative input guided by the material parameters~\cite{coropceanu2007charge}.

For a gapped independent-electron state, the molecular Berry curvature is
\begin{equation}
 b_i^z=-2\,\operatorname{Im}
 \sum_{n\in\mathrm{occ},\,m\in\mathrm{unocc}}
 \frac{\langle n|\partial_{u_i^x}H_{\rm el}|m\rangle
       \langle m|\partial_{u_i^y}H_{\rm el}|n\rangle}
      {(\varepsilon_m-\varepsilon_n)^2}.
 \label{eq:curvaturespectralmain}
\end{equation}
Here $H_{\rm el}$ is the one-electron Hamiltonian, $|n\rangle$ has energy $\varepsilon_n$, and $u_i^\alpha$ is a Cartesian ionic displacement.

I evaluate orbital magnetization separately from $b_i^z$.
Orbital circulation is evaluated in a projected L\"owdin frame constructed from occupied states and lower-potential $B$-site trial orbitals~\cite{lowdin1950non}.
The occupied-projector representation and its relation to this finite frame are given in SI Section~S1.
The plotted $M_{\rm IC}^{\rm res}=M^{\rm tot}-M_{\rm LC}$ subtracts the central-orbital local-circulation contribution from the total finite-flake magnetization.
With $e>0$ the elementary charge, the equivalent field $B_{{\rm eq},i}^z=(\hbar/e)b_i^z$ is a display scale, not an applied field or a calibration of $\epsilon_\star$.
More computational details are included in SI Sections~S1--S4.

\subsection{Phonon model and local OMP scattering}

The two-dimensional honeycomb model has two in-plane displacement coordinates per site and nearest-neighbor longitudinal and transverse springs.
I use $K_\parallel=21.998$ and $K_\perp=5.010$ eV\,\AA$^{-2}$, with $M_A=10.81$ and $M_B=14.00$ u~\cite{michel2009theory}.
Diagonalizing the mass-weighted dynamical matrix gives four branches, with $E_\nu=\hbar\sqrt{\lambda_\nu}$ for eigenvalue $\lambda_\nu$ in consistent units.
Initial, final, and virtual states in the transport kernel are restricted to the two acoustic branches, which is an approximation to describe the low-temperature behavior.

Each phonon-state index includes momentum and branch; $i$ and $f$ denote initial and final states.
Let $E_i>0$ be the phonon energy and $z_{\mu i}$ its two-component mass-weighted Cartesian polarization on sublattice $\mu$, normalized by $\sum_\mu z_{\mu i}^\dagger z_{\mu i}=1$.
Using the dimensionless reduced momenta $k_1,k_2$ (SI Section~S6) and the Bloch convention of SI Section~S7, define $\chi_A=1$, $\chi_B=\exp[i(k_1+k_2)/3]$, and $\widetilde z_{\mu i}=\chi_\mu z_{\mu i}$.
Substituting Eq.~\eqref{eq:Raman} with $\epsilon_\mu^{\rm tr}=\epsilon_\star s_\mu$ in place of $\epsilon_i^{\rm BO}$, one obtains
\begin{equation}
 U_{fi}=-\frac12\sum_{\mu=A,B}\epsilon_\mu^{\rm tr}
 \left(\sqrt{\frac{E_f}{E_i}}+
       \sqrt{\frac{E_i}{E_f}}\right)
 \widetilde z_{\mu f}^\dagger\sigma_y\widetilde z_{\mu i},
\end{equation}
where $\sigma_y$ is the Pauli matrix acting in Cartesian displacement space.
The A/B amplitudes add coherently within one scattering center; $s_A=0.86$ and $s_B=1.14$ are the normalized curvature weights defined above.
The vertex $U_{fi}$ has units of energy.
Independent dilute center positions are averaged before applying the mean-power approximation
$\langle\prod_{j=1}^{n}\epsilon_{a\mu_j}^{\rm tr}\rangle_\epsilon\simeq\prod_{j=1}^{n}\epsilon_{\mu_j}^{\rm tr}$ for $n=2,3$, where $a$ labels a center and $\mu_j=A,B$.
Here, $\epsilon_{a\mu}^{\rm tr}$ is the coupling on sublattice $\mu$ of center $a$, and $\langle\cdots\rangle_\epsilon$ averages over the local-coefficient distribution. SI Section~S6 describes the phonon model, and SI Section~S7 gives the vertex and the approximation for the disorder moments.

\subsection{Born scattering and Boltzmann transport}

For concentration $c_{\rm imp}$ per primitive cell, write $W^{(n)}=c_{\rm imp}\hat W^{(n)}+O(c_{\rm imp}^2)$, $n=2,3$~\cite{sinitsyn2007skew}.
On a uniform $N_k$-point mesh, the concentration-normalized Born scattering rates are
\begin{align}
 \hat W_{fi}^{(2)}
 &=\frac{2\pi}{\hbar}q_f |U_{fi}|^2
 \delta(E_f-E_i),
 \label{eq:W2}\\
 \hat W_{fi}^{(3)}
 &=\frac{4\pi}{\hbar}q_f\,
 \mathrm{Re}\!\left[
 U_{fi}^*\sum_vq_v
 \frac{U_{fv}U_{vi}}{E_i-E_v+i\eta}\right]
 \delta(E_f-E_i),
 \qquad q_f=q_v=\frac1{N_k}.
 \label{eq:W3}
\end{align}
The index $v$ labels virtual acoustic states.
The branch sums are included in the state indices.
I set $\hat W_{ii}^{(2)}=\hat W_{ii}^{(3)}=0$ to exclude forward scattering.
The sign reversal of $\epsilon_\mu^{\rm tr}$ leaves $\hat W^{(2)}$ even and makes $\hat W^{(3)}$ odd.
The skew scattering rate is $\hat W^{(3a)}=\tfrac12(\hat W^{(3)}-[\hat W^{(3)}]^{\mathsf T})$.
SI Section S8 includes more details.

I write the phonon occupation as $n_i=n_B(E_i,T)+g_i\partial_xT$, where $n_B$ is the Bose distribution, $g_i$ is the occupation response per temperature gradient, and $v_{i\alpha}$ is the group velocity along $\alpha=x,y$.
Define the concentration-normalized response $\hat g_i=c_{\rm imp}g_i$ and the driving term $d_i=v_{ix}\partial_Tn_B(E_i,T)$.
Expand $\hat g=\hat g^{(0)}+\hat g^{(1)}$ into the longitudinal solution and its leading skew-scattering correction.
For transition matrix $W$, whose entry $W_{fi}$ denotes a transition from $i$ to $f$, the collision matrix is
\begin{equation}
 C_{fi}=\delta_{fi}\sum_jW_{ji}-W_{fi}.
 \label{eq:collision}
\end{equation}
The leading-skew response satisfies
\begin{equation}
 \hat C_2\hat g^{(0)}=-d,
 \qquad
 \hat C_2\hat g^{(1)}=-\hat C_a\hat g^{(0)}.
 \label{eq:perturbative}
\end{equation}
Here $\hat C_2=C[\hat W^{(2)}]$ and $\hat C_a=C[\hat W^{(3a)}]$.
The longitudinal and transverse responses follow from $\hat g^{(0)}$ and $\hat g^{(1)}$, respectively.
The off-diagonal collision terms retain scattering into each phonon mode.
The constraint $\sum_i g_i=0$ fixes the number-conserving null mode.
The concentration-normalized conductivity is
\begin{align}
 \hat\kappa_{\alpha x}
 &\equiv\lim_{c_{\mathrm{imp}}\to0}c_{\mathrm{imp}}\kappa_{\alpha x}
 =-\frac{1}{N_k\Omega_3}\sum_i E_i v_{i\alpha}\hat g_i,
 \nonumber\\
 \Omega_3&=\Omega_2d,\qquad
 \Omega_2=\frac{\sqrt3}{2}a^2.
 \label{eq:kappa}
\end{align}
Here $\Omega_2$ is the primitive-cell area and $\Omega_3$ the reporting volume per cell.
Energies are converted to joules, $d=5$ \AA{} is the reporting thickness, and the heat-current density $j_\alpha^Q=-\kappa_{\alpha x}\partial_xT$ fixes the sign convention.
The odd component is half the conductivity difference between opposite $\epsilon_\mu^{\rm tr}$.
With OMP scattering as the only relaxation channel, $\kappa_{\alpha x}=\hat\kappa_{\alpha x}/c_{\rm imp}+O(1)$; the Hall angle is independent of $c_{\rm imp}$ at fixed coupling and of $d$. SI Section~S9 includes more details.

To compute Born scattering rates, I relate the impurity linewidth to the population lifetime through $\Gamma_i^{\mathrm{imp}}=\hbar/\tau_{\mathrm{pop},i}$ using Eq.~\eqref{eq:linewidthcalibrationmain}. I use $0.0119$ meV as an illustrative target for the unweighted median excess linewidth across both acoustic branches in the $8$–$12$ meV window, which is guided by the material parameters. At $c_{\mathrm{imp}}=0.01$ per primitive cell, I assume that OMP scattering centers dominate and obtain $|\epsilon_\star|\approx 10.0$ meV. Figure~\ref{fig:transport} uses the rounded value $\epsilon_\star=10$ meV.

I calculate both the longitudinal and transverse thermal conductivities from OMP scattering alone over 5–120 K, using a \(50\times50\) momentum grid and a numerical energy resolution of \(\eta=3\) meV. I obtain the group velocities from centered differences of the phonon energies and average the elastic scattering rates over shifts of the energy-bin boundaries (SI Sections~S6 and S8). I test numerical convergence at fixed \(\epsilon_\star\) (SI Section~S10).

\section*{AI usage}

During the preparation of this work, the author used ChatGPT and Codex (OpenAI) to assist with manuscript drafting and language editing, numerical-code development and debugging, checks of calculations and data consistency, and preparation of submission materials. The author takes full responsibility for the scientific content, numerical results, and final manuscript.

\backmatter
\section*{Data availability}
The numerical data supporting the findings of this study are provided in Supplementary Data 1.
\section*{Code availability}
The custom MATLAB code used for the numerical calculations, together with reproduction instructions and validation scripts, is provided in the accompanying \texttt{Research\_code.zip} archive.
\section*{Acknowledgements}
I thank Naoto Nagaosa and Junha Kang for fruitful discussions. T.O. was supported by the Basic Science Research Program through the National Research Foundation of Korea (NRF), funded by the Ministry of Education (RS-2021-NR060140). T.O. was also supported by the Pioneer Project for Future-Oriented Convergence Technology (Challenge Type; RS-2024-00416976) and the Basic Science Research Program (RS-2025-16065011 and RS-2026-25470523) through the National Research Foundation of Korea (NRF), funded by the Ministry of Science and ICT of the Republic of Korea. This work was also supported by the Nano \& Material Technology Development Program through the NRF, funded by the Ministry of Science and ICT (RS-2026-25607606), and by the G-LAMP Program of the NRF, funded by the Ministry of Education (RS-2025-25441317).
The funders had no role in the design of the study; the performance of the calculations; the analysis or interpretation of the results; the writing of the manuscript; or the decision to submit the manuscript for publication.
\section*{Author contributions}
T.O. conceived the study, developed the model, performed the calculations, analyzed the results, and wrote the manuscript.
\section*{Competing interests}
The author declares no competing interests.


\begin{thebibliography}{10}
\expandafter\ifx\csname url\endcsname\relax
  \def\url#1{\burl{#1}}\fi
\expandafter\ifx\csname urlprefix\endcsname\relax\def\urlprefix{URL }\fi
\providecommand{\bibinfo}[2]{#2}
\providecommand{\eprint}[2][]{\url{#2}}
\providecommand{\doi}[1]{\url{https://doi.org/#1}}
\bibcommenthead

\bibitem{strohm2005phenomenological}
\bibinfo{author}{Strohm, C.}, \bibinfo{author}{Rikken, G.} \&
  \bibinfo{author}{Wyder, P.}
\newblock \bibinfo{title}{Phenomenological evidence for the phonon {Hall}
  effect}.
\newblock \emph{\bibinfo{journal}{Physical review letters}}
  \textbf{\bibinfo{volume}{95}}, \bibinfo{pages}{155901}
  (\bibinfo{year}{2005}).

\bibitem{inyushkin2007phonon}
\bibinfo{author}{Inyushkin, A.~V.} \& \bibinfo{author}{Taldenkov, A.~N.}
\newblock \bibinfo{title}{On the phonon {Hall} effect in a paramagnetic
  dielectric}.
\newblock \emph{\bibinfo{journal}{Jetp Letters}} \textbf{\bibinfo{volume}{86}},
  \bibinfo{pages}{379--382} (\bibinfo{year}{2007}).

\bibitem{onose2010observation}
\bibinfo{author}{Onose, Y.} \emph{et~al.}
\newblock \bibinfo{title}{Observation of the magnon {Hall} effect}.
\newblock \emph{\bibinfo{journal}{Science}} \textbf{\bibinfo{volume}{329}},
  \bibinfo{pages}{297--299} (\bibinfo{year}{2010}).

\bibitem{mori2014origin}
\bibinfo{author}{Mori, M.}, \bibinfo{author}{Spencer-Smith, A.},
  \bibinfo{author}{Sushkov, O.~P.} \& \bibinfo{author}{Maekawa, S.}
\newblock \bibinfo{title}{Origin of the phonon {Hall} effect in rare-earth
  garnets}.
\newblock \emph{\bibinfo{journal}{Physical review letters}}
  \textbf{\bibinfo{volume}{113}}, \bibinfo{pages}{265901}
  (\bibinfo{year}{2014}).

\bibitem{hirschberger2015large}
\bibinfo{author}{Hirschberger, M.}, \bibinfo{author}{Krizan, J.~W.},
  \bibinfo{author}{Cava, R.} \& \bibinfo{author}{Ong, N.}
\newblock \bibinfo{title}{Large thermal {Hall} conductivity of neutral spin
  excitations in a frustrated quantum magnet}.
\newblock \emph{\bibinfo{journal}{Science}} \textbf{\bibinfo{volume}{348}},
  \bibinfo{pages}{106--109} (\bibinfo{year}{2015}).

\bibitem{hirschberger2015thermal}
\bibinfo{author}{Hirschberger, M.}, \bibinfo{author}{Chisnell, R.},
  \bibinfo{author}{Lee, Y.~S.} \& \bibinfo{author}{Ong, N.~P.}
\newblock \bibinfo{title}{Thermal {Hall} effect of spin excitations in a kagome
  magnet}.
\newblock \emph{\bibinfo{journal}{Physical review letters}}
  \textbf{\bibinfo{volume}{115}}, \bibinfo{pages}{106603}
  (\bibinfo{year}{2015}).

\bibitem{ideue2017giant}
\bibinfo{author}{Ideue, T.}, \bibinfo{author}{Kurumaji, T.},
  \bibinfo{author}{Ishiwata, S.} \& \bibinfo{author}{Tokura, Y.}
\newblock \bibinfo{title}{Giant thermal {Hall} effect in multiferroics}.
\newblock \emph{\bibinfo{journal}{Nature materials}}
  \textbf{\bibinfo{volume}{16}}, \bibinfo{pages}{797--802}
  (\bibinfo{year}{2017}).

\bibitem{zhang2021anomalous}
\bibinfo{author}{Zhang, H.} \emph{et~al.}
\newblock \bibinfo{title}{Anomalous thermal {Hall} effect in an insulating {van
  der Waals} magnet}.
\newblock \emph{\bibinfo{journal}{Physical Review Letters}}
  \textbf{\bibinfo{volume}{127}}, \bibinfo{pages}{247202}
  (\bibinfo{year}{2021}).

\bibitem{akazawa2022topological}
\bibinfo{author}{Akazawa, M.} \emph{et~al.}
\newblock \bibinfo{title}{Topological thermal {Hall} effect of magnons in
  magnetic skyrmion lattice}.
\newblock \emph{\bibinfo{journal}{Physical Review Research}}
  \textbf{\bibinfo{volume}{4}}, \bibinfo{pages}{043085} (\bibinfo{year}{2022}).

\bibitem{kim2024thermal}
\bibinfo{author}{Kim, H.-L.} \emph{et~al.}
\newblock \bibinfo{title}{Thermal {Hall} effects due to topological spin
  fluctuations in {{YMnO$_3$}}}.
\newblock \emph{\bibinfo{journal}{Nature Communications}}
  \textbf{\bibinfo{volume}{15}}, \bibinfo{pages}{243} (\bibinfo{year}{2024}).

\bibitem{kasahara2018majorana}
\bibinfo{author}{Kasahara, Y.} \emph{et~al.}
\newblock \bibinfo{title}{{Majorana} quantization and half-integer thermal
  quantum {Hall} effect in a {Kitaev} spin liquid}.
\newblock \emph{\bibinfo{journal}{Nature}} \textbf{\bibinfo{volume}{559}},
  \bibinfo{pages}{227--231} (\bibinfo{year}{2018}).

\bibitem{hentrich2018unusual}
\bibinfo{author}{Hentrich, R.} \emph{et~al.}
\newblock \bibinfo{title}{Unusual phonon heat transport in
  {$\alpha$-{RuCl}$_3$}: strong spin-phonon scattering and field-induced spin
  gap}.
\newblock \emph{\bibinfo{journal}{Physical review letters}}
  \textbf{\bibinfo{volume}{120}}, \bibinfo{pages}{117204}
  (\bibinfo{year}{2018}).

\bibitem{kasahara2018unusual}
\bibinfo{author}{Kasahara, Y.} \emph{et~al.}
\newblock \bibinfo{title}{Unusual thermal {Hall} effect in a {Kitaev} spin
  liquid candidate {$\alpha$-{RuCl}$_3$}}.
\newblock \emph{\bibinfo{journal}{Physical review letters}}
  \textbf{\bibinfo{volume}{120}}, \bibinfo{pages}{217205}
  (\bibinfo{year}{2018}).

\bibitem{hentrich2019large}
\bibinfo{author}{Hentrich, R.} \emph{et~al.}
\newblock \bibinfo{title}{Large thermal {Hall} effect in {$\alpha$-{RuCl}$_3$}:
  Evidence for heat transport by {Kitaev--Heisenberg} paramagnons}.
\newblock \emph{\bibinfo{journal}{Physical Review B}}
  \textbf{\bibinfo{volume}{99}}, \bibinfo{pages}{085136}
  (\bibinfo{year}{2019}).

\bibitem{lefranccois2022evidence}
\bibinfo{author}{Lefran{\c{c}}ois, {\'E}.} \emph{et~al.}
\newblock \bibinfo{title}{Evidence of a phonon {Hall} effect in the {Kitaev}
  spin liquid candidate {$\alpha$-{RuCl}$_3$}}.
\newblock \emph{\bibinfo{journal}{Physical Review X}}
  \textbf{\bibinfo{volume}{12}}, \bibinfo{pages}{021025}
  (\bibinfo{year}{2022}).

\bibitem{bruin2022robustness}
\bibinfo{author}{Bruin, J.} \emph{et~al.}
\newblock \bibinfo{title}{Robustness of the thermal {Hall} effect close to
  half-quantization in {$\alpha$-{RuCl}$_3$}}.
\newblock \emph{\bibinfo{journal}{Nature Physics}}
  \textbf{\bibinfo{volume}{18}}, \bibinfo{pages}{401--405}
  (\bibinfo{year}{2022}).

\bibitem{chen2024planar}
\bibinfo{author}{Chen, L.} \emph{et~al.}
\newblock \bibinfo{title}{Planar thermal {Hall} effect from phonons in a
  {Kitaev} candidate material}.
\newblock \emph{\bibinfo{journal}{Nature Communications}}
  \textbf{\bibinfo{volume}{15}}, \bibinfo{pages}{3513} (\bibinfo{year}{2024}).

\bibitem{grissonnanche2019giant}
\bibinfo{author}{Grissonnanche, G.} \emph{et~al.}
\newblock \bibinfo{title}{Giant thermal {Hall} conductivity in the pseudogap
  phase of cuprate superconductors}.
\newblock \emph{\bibinfo{journal}{Nature}} \textbf{\bibinfo{volume}{571}},
  \bibinfo{pages}{376--380} (\bibinfo{year}{2019}).

\bibitem{boulanger2020thermal}
\bibinfo{author}{Boulanger, M.-E.} \emph{et~al.}
\newblock \bibinfo{title}{Thermal {Hall} conductivity in the cuprate {Mott}
  insulators {{Nd$_2$CuO$_4$}} and {{Sr$_2$CuO$_2$Cl$_2$}}}.
\newblock \emph{\bibinfo{journal}{Nature communications}}
  \textbf{\bibinfo{volume}{11}}, \bibinfo{pages}{5325} (\bibinfo{year}{2020}).

\bibitem{katsura2010theory}
\bibinfo{author}{Katsura, H.}, \bibinfo{author}{Nagaosa, N.} \&
  \bibinfo{author}{Lee, P.~A.}
\newblock \bibinfo{title}{Theory of the thermal {Hall} effect in quantum
  magnets}.
\newblock \emph{\bibinfo{journal}{Physical review letters}}
  \textbf{\bibinfo{volume}{104}}, \bibinfo{pages}{066403}
  (\bibinfo{year}{2010}).

\bibitem{matsumoto2011rotational}
\bibinfo{author}{Matsumoto, R.} \& \bibinfo{author}{Murakami, S.}
\newblock \bibinfo{title}{Rotational motion of magnons and the thermal {Hall}
  effect}.
\newblock \emph{\bibinfo{journal}{Physical Review B—Condensed Matter and
  Materials Physics}} \textbf{\bibinfo{volume}{84}}, \bibinfo{pages}{184406}
  (\bibinfo{year}{2011}).

\bibitem{owerre2016first}
\bibinfo{author}{Owerre, S.}
\newblock \bibinfo{title}{A first theoretical realization of honeycomb
  topological magnon insulator}.
\newblock \emph{\bibinfo{journal}{Journal of Physics: Condensed Matter}}
  \textbf{\bibinfo{volume}{28}}, \bibinfo{pages}{386001}
  (\bibinfo{year}{2016}).

\bibitem{owerre2017topological}
\bibinfo{author}{Owerre, S.}
\newblock \bibinfo{title}{Topological thermal {Hall} effect in frustrated
  kagome antiferromagnets}.
\newblock \emph{\bibinfo{journal}{Physical Review B}}
  \textbf{\bibinfo{volume}{95}}, \bibinfo{pages}{014422}
  (\bibinfo{year}{2017}).

\bibitem{zhang2019thermal}
\bibinfo{author}{Zhang, X.}, \bibinfo{author}{Zhang, Y.},
  \bibinfo{author}{Okamoto, S.} \& \bibinfo{author}{Xiao, D.}
\newblock \bibinfo{title}{Thermal {Hall} effect induced by magnon-phonon
  interactions}.
\newblock \emph{\bibinfo{journal}{Physical review letters}}
  \textbf{\bibinfo{volume}{123}}, \bibinfo{pages}{167202}
  (\bibinfo{year}{2019}).

\bibitem{zhang2021topological}
\bibinfo{author}{Zhang, E.~Z.}, \bibinfo{author}{Chern, L.~E.} \&
  \bibinfo{author}{Kim, Y.~B.}
\newblock \bibinfo{title}{Topological magnons for thermal {Hall} transport in
  frustrated magnets with bond-dependent interactions}.
\newblock \emph{\bibinfo{journal}{Physical Review B}}
  \textbf{\bibinfo{volume}{103}}, \bibinfo{pages}{174402}
  (\bibinfo{year}{2021}).

\bibitem{zhang2024thermal}
\bibinfo{author}{Zhang, X.-T.}, \bibinfo{author}{Gao, Y.~H.} \&
  \bibinfo{author}{Chen, G.}
\newblock \bibinfo{title}{Thermal {Hall} effects in quantum magnets}.
\newblock \emph{\bibinfo{journal}{Physics Reports}}
  \textbf{\bibinfo{volume}{1070}}, \bibinfo{pages}{1--59}
  (\bibinfo{year}{2024}).

\bibitem{sheng2006theory}
\bibinfo{author}{Sheng, L.}, \bibinfo{author}{Sheng, D.} \&
  \bibinfo{author}{Ting, C.}
\newblock \bibinfo{title}{Theory of the phonon {Hall} effect in paramagnetic
  dielectrics}.
\newblock \emph{\bibinfo{journal}{Physical review letters}}
  \textbf{\bibinfo{volume}{96}}, \bibinfo{pages}{155901}
  (\bibinfo{year}{2006}).

\bibitem{kagan2008anomalous}
\bibinfo{author}{Kagan, Y.} \& \bibinfo{author}{Maksimov, L.}
\newblock \bibinfo{title}{Anomalous {Hall} effect for the phonon heat
  conductivity in paramagnetic dielectrics}.
\newblock \emph{\bibinfo{journal}{Physical review letters}}
  \textbf{\bibinfo{volume}{100}}, \bibinfo{pages}{145902}
  (\bibinfo{year}{2008}).

\bibitem{wang2009phonon}
\bibinfo{author}{Wang, J.-S.} \& \bibinfo{author}{Zhang, L.}
\newblock \bibinfo{title}{Phonon {Hall} thermal conductivity from the
  {Green--Kubo} formula}.
\newblock \emph{\bibinfo{journal}{Physical Review B—Condensed Matter and
  Materials Physics}} \textbf{\bibinfo{volume}{80}}, \bibinfo{pages}{012301}
  (\bibinfo{year}{2009}).

\bibitem{zhang2010topological}
\bibinfo{author}{Zhang, L.}, \bibinfo{author}{Ren, J.}, \bibinfo{author}{Wang,
  J.-S.} \& \bibinfo{author}{Li, B.}
\newblock \bibinfo{title}{Topological nature of the phonon {Hall} effect}.
\newblock \emph{\bibinfo{journal}{Physical review letters}}
  \textbf{\bibinfo{volume}{105}}, \bibinfo{pages}{225901}
  (\bibinfo{year}{2010}).

\bibitem{agarwalla2011phonon}
\bibinfo{author}{Agarwalla, B.~K.}, \bibinfo{author}{Zhang, L.},
  \bibinfo{author}{Wang, J.-S.} \& \bibinfo{author}{Li, B.}
\newblock \bibinfo{title}{Phonon {Hall} effect in ionic crystals in the
  presence of static magnetic field}.
\newblock \emph{\bibinfo{journal}{The European Physical Journal B}}
  \textbf{\bibinfo{volume}{81}}, \bibinfo{pages}{197--202}
  (\bibinfo{year}{2011}).

\bibitem{qin2012berry}
\bibinfo{author}{Qin, T.}, \bibinfo{author}{Zhou, J.} \& \bibinfo{author}{Shi,
  J.}
\newblock \bibinfo{title}{{Berry} curvature and the phonon {Hall} effect}.
\newblock \emph{\bibinfo{journal}{Physical Review B—Condensed Matter and
  Materials Physics}} \textbf{\bibinfo{volume}{86}}, \bibinfo{pages}{104305}
  (\bibinfo{year}{2012}).

\bibitem{saito2019berry}
\bibinfo{author}{Saito, T.}, \bibinfo{author}{Misaki, K.},
  \bibinfo{author}{Ishizuka, H.} \& \bibinfo{author}{Nagaosa, N.}
\newblock \bibinfo{title}{{Berry} phase of phonons and thermal {Hall} effect in
  nonmagnetic insulators}.
\newblock \emph{\bibinfo{journal}{Physical Review Letters}}
  \textbf{\bibinfo{volume}{123}}, \bibinfo{pages}{255901}
  (\bibinfo{year}{2019}).

\bibitem{oh2025phonon}
\bibinfo{author}{Oh, T.} \& \bibinfo{author}{Nagaosa, N.}
\newblock \bibinfo{title}{Phonon thermal {Hall} effect in {Mott} insulators via
  skew scattering by the scalar spin chirality}.
\newblock \emph{\bibinfo{journal}{Physical Review X}}
  \textbf{\bibinfo{volume}{15}}, \bibinfo{pages}{011036}
  (\bibinfo{year}{2025}).

\bibitem{oh2025spin}
\bibinfo{author}{Oh, T.} \& \bibinfo{author}{Nagaosa, N.}
\newblock \bibinfo{title}{Spin-phonon coupling and thermal {Hall} effect in the
  {Kitaev} model}.
\newblock \emph{\bibinfo{journal}{Physical Review B}}
  \textbf{\bibinfo{volume}{112}}, \bibinfo{pages}{L081104}
  (\bibinfo{year}{2025}).

\bibitem{mangeolle2022phonon}
\bibinfo{author}{Mangeolle, L.}, \bibinfo{author}{Balents, L.} \&
  \bibinfo{author}{Savary, L.}
\newblock \bibinfo{title}{Phonon thermal {Hall} conductivity from scattering
  with collective fluctuations}.
\newblock \emph{\bibinfo{journal}{Physical Review X}}
  \textbf{\bibinfo{volume}{12}}, \bibinfo{pages}{041031}
  (\bibinfo{year}{2022}).

\bibitem{sun2022large}
\bibinfo{author}{Sun, X.-Q.}, \bibinfo{author}{Chen, J.-Y.} \&
  \bibinfo{author}{Kivelson, S.~A.}
\newblock \bibinfo{title}{Large extrinsic phonon thermal {Hall} effect from
  resonant scattering}.
\newblock \emph{\bibinfo{journal}{Physical Review B}}
  \textbf{\bibinfo{volume}{106}}, \bibinfo{pages}{144111}
  (\bibinfo{year}{2022}).

\bibitem{li2020phonon}
\bibinfo{author}{Li, X.}, \bibinfo{author}{Fauqu{\'e}, B.},
  \bibinfo{author}{Zhu, Z.} \& \bibinfo{author}{Behnia, K.}
\newblock \bibinfo{title}{Phonon thermal {Hall} effect in strontium titanate}.
\newblock \emph{\bibinfo{journal}{Physical review letters}}
  \textbf{\bibinfo{volume}{124}}, \bibinfo{pages}{105901}
  (\bibinfo{year}{2020}).

\bibitem{sim2021sizable}
\bibinfo{author}{Sim, S.} \emph{et~al.}
\newblock \bibinfo{title}{Sizable suppression of thermal {Hall} effect upon
  isotopic substitution in {{SrTiO$_3$}}}.
\newblock \emph{\bibinfo{journal}{Physical Review Letters}}
  \textbf{\bibinfo{volume}{126}}, \bibinfo{pages}{015901}
  (\bibinfo{year}{2021}).

\bibitem{sharma2024phonon}
\bibinfo{author}{Sharma, R.}, \bibinfo{author}{Bagchi, M.},
  \bibinfo{author}{Wang, Y.}, \bibinfo{author}{Ando, Y.} \&
  \bibinfo{author}{Lorenz, T.}
\newblock \bibinfo{title}{Phonon thermal {Hall} effect in charge-compensated
  topological insulators}.
\newblock \emph{\bibinfo{journal}{Physical Review B}}
  \textbf{\bibinfo{volume}{109}}, \bibinfo{pages}{104304}
  (\bibinfo{year}{2024}).

\bibitem{sharma2024phonon2}
\bibinfo{author}{Sharma, R.}, \bibinfo{author}{Valldor, M.} \&
  \bibinfo{author}{Lorenz, T.}
\newblock \bibinfo{title}{Phonon thermal {Hall} effect in nonmagnetic
  {{Y$_2$Ti$_2$O$_7$}}}.
\newblock \emph{\bibinfo{journal}{Physical Review B}}
  \textbf{\bibinfo{volume}{110}}, \bibinfo{pages}{L100301}
  (\bibinfo{year}{2024}).

\bibitem{li2023phonon}
\bibinfo{author}{Li, X.} \emph{et~al.}
\newblock \bibinfo{title}{The phonon thermal {Hall} angle in black phosphorus}.
\newblock \emph{\bibinfo{journal}{Nature Communications}}
  \textbf{\bibinfo{volume}{14}}, \bibinfo{pages}{1027} (\bibinfo{year}{2023}).

\bibitem{jin2024discovery}
\bibinfo{author}{Jin, X.} \emph{et~al.}
\newblock \bibinfo{title}{Discovery of universal phonon thermal {Hall} effect
  in crystals}.
\newblock \emph{\bibinfo{journal}{Physical Review Letters}}
  \textbf{\bibinfo{volume}{135}}, \bibinfo{pages}{196302}
  (\bibinfo{year}{2025}).

\bibitem{thonhauser2005orbital}
\bibinfo{author}{Thonhauser, T.}, \bibinfo{author}{Ceresoli, D.},
  \bibinfo{author}{Vanderbilt, D.} \& \bibinfo{author}{Resta, R.}
\newblock \bibinfo{title}{Orbital magnetization in periodic insulators}.
\newblock \emph{\bibinfo{journal}{Physical review letters}}
  \textbf{\bibinfo{volume}{95}}, \bibinfo{pages}{137205}
  (\bibinfo{year}{2005}).

\bibitem{xiao2005berry}
\bibinfo{author}{Xiao, D.}, \bibinfo{author}{Shi, J.} \& \bibinfo{author}{Niu,
  Q.}
\newblock \bibinfo{title}{{Berry} phase correction to electron density of
  states in solids}.
\newblock \emph{\bibinfo{journal}{Physical review letters}}
  \textbf{\bibinfo{volume}{95}}, \bibinfo{pages}{137204}
  (\bibinfo{year}{2005}).

\bibitem{ceresoli2006orbital}
\bibinfo{author}{Ceresoli, D.}, \bibinfo{author}{Thonhauser, T.},
  \bibinfo{author}{Vanderbilt, D.} \& \bibinfo{author}{Resta, R.}
\newblock \bibinfo{title}{Orbital magnetization in crystalline solids:
  Multi-band insulators, {Chern} insulators, and metals}.
\newblock \emph{\bibinfo{journal}{Physical Review B—Condensed Matter and
  Materials Physics}} \textbf{\bibinfo{volume}{74}}, \bibinfo{pages}{024408}
  (\bibinfo{year}{2006}).

\bibitem{shi2007quantum}
\bibinfo{author}{Shi, J.}, \bibinfo{author}{Vignale, G.},
  \bibinfo{author}{Xiao, D.} \& \bibinfo{author}{Niu, Q.}
\newblock \bibinfo{title}{Quantum theory of orbital magnetization and its
  generalization to interacting systems}.
\newblock \emph{\bibinfo{journal}{Physical review letters}}
  \textbf{\bibinfo{volume}{99}}, \bibinfo{pages}{197202}
  (\bibinfo{year}{2007}).

\bibitem{xiao2010berry}
\bibinfo{author}{Xiao, D.}, \bibinfo{author}{Chang, M.-C.} \&
  \bibinfo{author}{Niu, Q.}
\newblock \bibinfo{title}{{Berry} phase effects on electronic properties}.
\newblock \emph{\bibinfo{journal}{Reviews of modern physics}}
  \textbf{\bibinfo{volume}{82}}, \bibinfo{pages}{1959--2007}
  (\bibinfo{year}{2010}).

\bibitem{thonhauser2011theory}
\bibinfo{author}{Thonhauser, T.}
\newblock \bibinfo{title}{Theory of orbital magnetization in solids}.
\newblock \emph{\bibinfo{journal}{International Journal of Modern Physics B}}
  \textbf{\bibinfo{volume}{25}}, \bibinfo{pages}{1429--1458}
  (\bibinfo{year}{2011}).

\bibitem{aryasetiawan2019modern}
\bibinfo{author}{Aryasetiawan, F.} \& \bibinfo{author}{Karlsson, K.}
\newblock \bibinfo{title}{Modern theory of orbital magnetic moment in solids}.
\newblock \emph{\bibinfo{journal}{Journal of Physics and Chemistry of Solids}}
  \textbf{\bibinfo{volume}{128}}, \bibinfo{pages}{87--108}
  (\bibinfo{year}{2019}).

\bibitem{haldane1988model}
\bibinfo{author}{Haldane, F. D.~M.}
\newblock \bibinfo{title}{Model for a quantum {Hall} effect without {Landau}
  levels: Condensed-matter realization of the ``parity anomaly''}.
\newblock \emph{\bibinfo{journal}{Physical review letters}}
  \textbf{\bibinfo{volume}{61}}, \bibinfo{pages}{2015} (\bibinfo{year}{1988}).

\bibitem{thonhauser2006insulator}
\bibinfo{author}{Thonhauser, T.} \& \bibinfo{author}{Vanderbilt, D.}
\newblock \bibinfo{title}{Insulator/{Chern}-insulator transition in the
  {Haldane} model}.
\newblock \emph{\bibinfo{journal}{Physical Review B—Condensed Matter and
  Materials Physics}} \textbf{\bibinfo{volume}{74}}, \bibinfo{pages}{235111}
  (\bibinfo{year}{2006}).

\bibitem{lovesey2002orbital}
\bibinfo{author}{Lovesey, S.~W.}, \bibinfo{author}{Knight, K.} \&
  \bibinfo{author}{Sivia, D.}
\newblock \bibinfo{title}{Orbital magnetization of a {Mott} insulator,
  {{V$_2$O$_3$}}, revealed by resonant x-ray {Bragg} diffraction}.
\newblock \emph{\bibinfo{journal}{Physical Review B}}
  \textbf{\bibinfo{volume}{65}}, \bibinfo{pages}{224402}
  (\bibinfo{year}{2002}).

\bibitem{braude2006orbital}
\bibinfo{author}{Braude, V.} \& \bibinfo{author}{Sonin, E.}
\newblock \bibinfo{title}{Orbital magnetic dynamics in chiral p-wave
  superconductors}.
\newblock \emph{\bibinfo{journal}{Physical Review B—Condensed Matter and
  Materials Physics}} \textbf{\bibinfo{volume}{74}}, \bibinfo{pages}{064501}
  (\bibinfo{year}{2006}).

\bibitem{holmvall2023enhanced}
\bibinfo{author}{Holmvall, P.} \& \bibinfo{author}{Black-Schaffer, A.~M.}
\newblock \bibinfo{title}{Enhanced chiral edge currents and orbital magnetic
  moment in chiral d-wave superconductors from mesoscopic finite-size effects}.
\newblock \emph{\bibinfo{journal}{Physical Review B}}
  \textbf{\bibinfo{volume}{108}}, \bibinfo{pages}{174505}
  (\bibinfo{year}{2023}).

\bibitem{meyer1961experimental}
\bibinfo{author}{Meyer, A.} \& \bibinfo{author}{Asch, G.}
\newblock \bibinfo{title}{Experimental $g'$ and $g$ values of {Fe}, {Co}, {Ni},
  and their alloys}.
\newblock \emph{\bibinfo{journal}{Journal of Applied Physics}}
  \textbf{\bibinfo{volume}{32}}, \bibinfo{pages}{S330--S333}
  (\bibinfo{year}{1961}).

\bibitem{eriksson1990orbital}
\bibinfo{author}{Eriksson, O.}, \bibinfo{author}{Johansson, B.} \&
  \bibinfo{author}{Brooks, M.}
\newblock \bibinfo{title}{Orbital magnetism in the itinerant ferromagnet
  {{NpOs$_2$}}}.
\newblock \emph{\bibinfo{journal}{Physical Review B}}
  \textbf{\bibinfo{volume}{41}}, \bibinfo{pages}{9095} (\bibinfo{year}{1990}).

\bibitem{hiess2001spin}
\bibinfo{author}{Hiess, A.} \emph{et~al.}
\newblock \bibinfo{title}{Spin and orbital moments in itinerant magnets}.
\newblock \emph{\bibinfo{journal}{Europhysics letters}}
  \textbf{\bibinfo{volume}{55}}, \bibinfo{pages}{267} (\bibinfo{year}{2001}).

\bibitem{solovyev2005orbital}
\bibinfo{author}{Solovyev, I.}
\newblock \bibinfo{title}{Orbital polarization in itinerant magnets}.
\newblock \emph{\bibinfo{journal}{Physical review letters}}
  \textbf{\bibinfo{volume}{95}}, \bibinfo{pages}{267205}
  (\bibinfo{year}{2005}).

\bibitem{ceresoli2010first}
\bibinfo{author}{Ceresoli, D.}, \bibinfo{author}{Gerstmann, U.},
  \bibinfo{author}{Seitsonen, A.~P.} \& \bibinfo{author}{Mauri, F.}
\newblock \bibinfo{title}{First-principles theory of orbital magnetization}.
\newblock \emph{\bibinfo{journal}{Physical Review B—Condensed Matter and
  Materials Physics}} \textbf{\bibinfo{volume}{81}}, \bibinfo{pages}{060409}
  (\bibinfo{year}{2010}).

\bibitem{ovesen2024orbital}
\bibinfo{author}{Ovesen, M.} \& \bibinfo{author}{Olsen, T.}
\newblock \bibinfo{title}{Orbital magnetization in two-dimensional materials
  from high-throughput computational screening}.
\newblock \emph{\bibinfo{journal}{2D Materials}} \textbf{\bibinfo{volume}{11}},
  \bibinfo{pages}{045010} (\bibinfo{year}{2024}).

\bibitem{cheng2009orbital}
\bibinfo{author}{Cheng, F.}, \bibinfo{author}{Zhi-Gang, W.},
  \bibinfo{author}{Shu-Shen, L.} \& \bibinfo{author}{Ping, Z.}
\newblock \bibinfo{title}{Orbital magnetization in semiconductors}.
\newblock \emph{\bibinfo{journal}{Chinese Physics B}}
  \textbf{\bibinfo{volume}{18}}, \bibinfo{pages}{5431} (\bibinfo{year}{2009}).

\bibitem{sliwa2014orbital}
\bibinfo{author}{{\'S}liwa, C.} \& \bibinfo{author}{Dietl, T.}
\newblock \bibinfo{title}{Orbital magnetization in dilute ferromagnetic
  semiconductors}.
\newblock \emph{\bibinfo{journal}{Physical Review B}}
  \textbf{\bibinfo{volume}{90}}, \bibinfo{pages}{045202}
  (\bibinfo{year}{2014}).

\bibitem{hasan2010colloquium}
\bibinfo{author}{Hasan, M.~Z.} \& \bibinfo{author}{Kane, C.~L.}
\newblock \bibinfo{title}{Colloquium: topological insulators}.
\newblock \emph{\bibinfo{journal}{Reviews of modern physics}}
  \textbf{\bibinfo{volume}{82}}, \bibinfo{pages}{3045--3067}
  (\bibinfo{year}{2010}).

\bibitem{michel2009theory}
\bibinfo{author}{Michel, K.} \& \bibinfo{author}{Verberck, B.}
\newblock \bibinfo{title}{Theory of elastic and piezoelectric effects in
  two-dimensional hexagonal boron nitride}.
\newblock \emph{\bibinfo{journal}{Physical Review B—Condensed Matter and
  Materials Physics}} \textbf{\bibinfo{volume}{80}}, \bibinfo{pages}{224301}
  (\bibinfo{year}{2009}).

\bibitem{kulda2001thermal}
\bibinfo{author}{Kulda, J.}, \bibinfo{author}{Farhi, E.} \&
  \bibinfo{author}{Zeyen, C. M.~E.}
\newblock \bibinfo{title}{Thermal variation of phonon frequency and line width
  in {Ge} studied by {TAS} spin-echo}.
\newblock \emph{\bibinfo{journal}{Physica B: Condensed Matter}}
  \textbf{\bibinfo{volume}{297}}, \bibinfo{pages}{37--39}
  (\bibinfo{year}{2001}).

\bibitem{chou2019long}
\bibinfo{author}{Chou, T.-H.} \emph{et~al.}
\newblock \bibinfo{title}{Long mean free paths of room-temperature {THz}
  acoustic phonons in a high thermal conductivity material}.
\newblock \emph{\bibinfo{journal}{Physical Review B}}
  \textbf{\bibinfo{volume}{100}}, \bibinfo{pages}{094302}
  (\bibinfo{year}{2019}).

\bibitem{tserkovnyak2015spin}
\bibinfo{author}{Tserkovnyak, Y.}, \bibinfo{author}{Pesin, D.~A.} \&
  \bibinfo{author}{Loss, D.}
\newblock \bibinfo{title}{Spin and orbital magnetic response on the surface of
  a topological insulator}.
\newblock \emph{\bibinfo{journal}{Physical Review B}}
  \textbf{\bibinfo{volume}{91}}, \bibinfo{pages}{041121}
  (\bibinfo{year}{2015}).

\bibitem{kim2017realizing}
\bibinfo{author}{Kim, H.-S.} \& \bibinfo{author}{Kee, H.-Y.}
\newblock \bibinfo{title}{Realizing {Haldane} model in {Fe}-based honeycomb
  ferromagnetic insulators}.
\newblock \emph{\bibinfo{journal}{npj Quantum Materials}}
  \textbf{\bibinfo{volume}{2}}, \bibinfo{pages}{20} (\bibinfo{year}{2017}).

\bibitem{urazhdin2025atomic}
\bibinfo{author}{Urazhdin, S.}
\newblock \bibinfo{title}{Atomic and interatomic orbital magnetization induced
  in {{SrTiO$_3$}} by chiral phonons}.
\newblock \emph{\bibinfo{journal}{Physical Review B}}
  \textbf{\bibinfo{volume}{111}}, \bibinfo{pages}{214435}
  (\bibinfo{year}{2025}).

\bibitem{kim2016unexpected}
\bibinfo{author}{Kim, H.~J.} \emph{et~al.}
\newblock \bibinfo{title}{Unexpected orbital magnetism in {Bi}-rich
  {{Bi$_2$Se$_3$}} nanoplatelets}.
\newblock \emph{\bibinfo{journal}{NPG Asia Materials}}
  \textbf{\bibinfo{volume}{8}}, \bibinfo{pages}{e271--e271}
  (\bibinfo{year}{2016}).

\bibitem{cullen2025giant}
\bibinfo{author}{Cullen, J.~H.}, \bibinfo{author}{Liu, H.} \&
  \bibinfo{author}{Culcer, D.}
\newblock \bibinfo{title}{Giant orbital {Hall} effect due to the bulk states of
  {3D} topological insulators}.
\newblock \emph{\bibinfo{journal}{npj Spintronics}}
  \textbf{\bibinfo{volume}{3}}, \bibinfo{pages}{22} (\bibinfo{year}{2025}).

\bibitem{gobel2024orbital}
\bibinfo{author}{G{\"o}bel, B.} \& \bibinfo{author}{Mertig, I.}
\newblock \bibinfo{title}{Orbital {Hall} effect accompanying quantum {Hall}
  effect: {Landau} levels cause orbital polarized edge currents}.
\newblock \emph{\bibinfo{journal}{Physical Review Letters}}
  \textbf{\bibinfo{volume}{133}}, \bibinfo{pages}{146301}
  (\bibinfo{year}{2024}).

\bibitem{matsumoto2025observation}
\bibinfo{author}{Matsumoto, R.} \emph{et~al.}
\newblock \bibinfo{title}{Observation of orbital {Hall} effect in {Si}}.
\newblock \emph{\bibinfo{journal}{arXiv preprint arXiv:2501.14237v3}}
  (\bibinfo{year}{2025}).

\bibitem{santos2024negative}
\bibinfo{author}{Santos, E.} \emph{et~al.}
\newblock \bibinfo{title}{Negative orbital {Hall} effect in germanium}.
\newblock \emph{\bibinfo{journal}{Physical Review Applied}}
  \textbf{\bibinfo{volume}{22}}, \bibinfo{pages}{064071}
  (\bibinfo{year}{2024}).

\bibitem{cysne2023ultrathin}
\bibinfo{author}{Cysne, T.~P.}, \bibinfo{author}{Costa, M.},
  \bibinfo{author}{Nardelli, M.~B.}, \bibinfo{author}{Muniz, R.} \&
  \bibinfo{author}{Rappoport, T.~G.}
\newblock \bibinfo{title}{Ultrathin films of black phosphorus as suitable
  platforms for unambiguous observation of the orbital {Hall} effect}.
\newblock \emph{\bibinfo{journal}{Physical Review B}}
  \textbf{\bibinfo{volume}{108}}, \bibinfo{pages}{165415}
  (\bibinfo{year}{2023}).

\bibitem{robinson2021large}
\bibinfo{author}{Robinson, R.~A.} \emph{et~al.}
\newblock \bibinfo{title}{Large violation of the {Wiedemann--Franz} law in
  {Heusler}, ferromagnetic, {Weyl} semimetal {{Co$_2$MnAl}}}.
\newblock \emph{\bibinfo{journal}{Journal of Physics D: Applied Physics}}
  \textbf{\bibinfo{volume}{54}}, \bibinfo{pages}{454001}
  (\bibinfo{year}{2021}).

\bibitem{park2020phonon}
\bibinfo{author}{Park, S.} \& \bibinfo{author}{Yang, B.-J.}
\newblock \bibinfo{title}{Phonon angular momentum {Hall} effect}.
\newblock \emph{\bibinfo{journal}{Nano Letters}} \textbf{\bibinfo{volume}{20}},
  \bibinfo{pages}{7694--7699} (\bibinfo{year}{2020}).

\bibitem{nagaosa2010anomalous}
\bibinfo{author}{Nagaosa, N.}, \bibinfo{author}{Sinova, J.},
  \bibinfo{author}{Onoda, S.}, \bibinfo{author}{MacDonald, A.~H.} \&
  \bibinfo{author}{Ong, N.~P.}
\newblock \bibinfo{title}{Anomalous {Hall} effect}.
\newblock \emph{\bibinfo{journal}{Reviews of modern physics}}
  \textbf{\bibinfo{volume}{82}}, \bibinfo{pages}{1539--1592}
  (\bibinfo{year}{2010}).

\bibitem{coropceanu2007charge}
\bibinfo{author}{Coropceanu, V.} \emph{et~al.}
\newblock \bibinfo{title}{Charge transport in organic semiconductors}.
\newblock \emph{\bibinfo{journal}{Chemical reviews}}
  \textbf{\bibinfo{volume}{107}}, \bibinfo{pages}{926--952}
  (\bibinfo{year}{2007}).

\bibitem{lowdin1950non}
\bibinfo{author}{L{\"o}wdin, P.-O.}
\newblock \bibinfo{title}{On the non-orthogonality problem connected with the
  use of atomic wave functions in the theory of molecules and crystals}.
\newblock \emph{\bibinfo{journal}{The Journal of Chemical Physics}}
  \textbf{\bibinfo{volume}{18}}, \bibinfo{pages}{365--375}
  (\bibinfo{year}{1950}).

\bibitem{sinitsyn2007skew}
\bibinfo{author}{Sinitsyn, N.~A.}, \bibinfo{author}{MacDonald, A.~H.},
  \bibinfo{author}{Jungwirth, T.}, \bibinfo{author}{Dugaev, V.~K.} \&
  \bibinfo{author}{Sinova, J.}
\newblock \bibinfo{title}{Anomalous {Hall} effect in a two-dimensional {Dirac}
  band: The link between the {Kubo--Streda} formula and the semiclassical
  {Boltzmann} equation approach}.
\newblock \emph{\bibinfo{journal}{Physical Review B}}
  \textbf{\bibinfo{volume}{75}}, \bibinfo{pages}{045315}
  (\bibinfo{year}{2007}).

\end{thebibliography}
\end{document}


\title{Supplementary Information\\Thermal Hall effect induced by phonon skew scattering via orbital magnetization}
\author{Taekoo Oh\\[2pt]\small Department of Physics, Soongsil University, Seoul 06978, Korea\\\small\texttt{taekoo.oh@ssu.ac.kr}}
\date{}
\maketitle

\tableofcontents\clearpage
\section{Finite-flake Haldane model, molecular Berry curvature, and Löwdin orbitals}
\label{sec:electronic}

The electronic reference model is the Haldane model~\cite{Haldane1988}:
\begin{equation}
 H=\left[\sum_{\langle ij\rangle}t_1 c_i^\dagger c_j
 +\sum_{\langle\!\langle ij\rangle\!\rangle}
 t_2 e^{\ii\varphi_{ij}}c_i^\dagger c_j\right]
 +\mathrm{H.c.}+\sum_i\xi_i c_i^\dagger c_i.
 \label{eq:Haldane}
\end{equation}
Here $c_i$ annihilates a spinless electron at site $i$. The two bond sums run over nearest and next-nearest neighbors, respectively, with one orientation per bond. For counterclockwise hopping from $j$ to $i$, $\varphi_{ij}=\varphi$, with $\varphi_{ji}=-\varphi_{ij}$ for the reverse hop. The onsite energies are chosen to be $\xi_A=-\xi_B=E_0>0$, without loss of generality.
I use $t_1=1$ eV, $t_2=t_1/4$, and a lattice constant $a=5$ \AA{} for a $16\times16$-cell open flake. With an $A$ site at the origin, the nearest-neighbor (NN) $B$ sites are at
\alg{
\bm \dt_1 = a(\f{1}{2},\f{\sqrt{3}}{6}), \qquad
\bm \dt_2 = a(-\f{1}{2},\f{\sqrt{3}}{6}), \qquad
\bm \dt_3 = a(0,-\f{1}{\sqrt{3}}).
}
The next-nearest-neighbor (NNN) sites are at
\begin{equation}
 \pm \bm a_1=\pm a(1,0),\qquad \pm \bm a_2=\pm a(\f{1}{2},\f{\sqrt3}{2}),\qquad \pm (\bm a_2-\bm a_1)=\pm a(-\f{1}{2},\f{\sqrt3}{2}),
 \label{eq:NNN-directions}
\end{equation}
around a bulk site.

I model the dependence of the hopping parameters on ionic displacements by a linear modulation of the three NN hoppings,
\begin{equation}
 \delta t_{ij}=g\,\hat{\bm d}_{ij}\!\cdot
 (\bm u_j-\bm u_i),\qquad
 g=0.1\ \mathrm{eV}\,\text{\AA}^{-1}.
 \label{eq:dtdu}
\end{equation}
Here, $\bm{\hat{d}}_{ij}$ is the unit vector along the original $ij$ bond.
The derivative in Eq.~\eqref{eq:dtdu} uses the Cartesian unit vector of each NN bond.
The value $g=0.1$ eV\,\AA$^{-1}$ is used as a conservative illustrative input; no material-specific calibration of this hopping derivative is claimed.

Next, let $P$ be the occupied-state projector of Eq.~\eqref{eq:Haldane}.  The
gauge-invariant molecular Berry curvature associated with two displacement
coordinates is
\begin{align}
 \mathcal F_{ab}
 &=+\ii\Tr\!\left(P[\partial_aP,\partial_bP]\right),
 \label{eq:Fprojector}\\
 &=-2\,\mathrm{Im}\sum_{o\in\mathrm{occ}}
 \sum_{u\in\mathrm{unocc}}
 \frac{\langle o|\partial_aH|u\rangle
       \langle u|\partial_bH|o\rangle}
 {(E_o-E_u)^2}.
 \label{eq:Fspectral}
\end{align}

The projected L\"owdin construction~\cite{Lowdin1950} describes localized electronic circulation in the finite flake.
Because $B$ is chosen to be the lower-potential sublattice, I take the trial columns
$T_B=(|1,B\rangle,\ldots,|N_{\rm cell},B\rangle)$, where $N_{\rm cell}$ is the number of primitive cells, and define the projected trials
$|\psi_{iB}\rangle=P|i,B\rangle$.  Collecting these columns gives
\begin{equation}
 X=PT_B,\qquad S=X^\dagger X,\qquad
 W=XS^{-1/2}.
 \label{eq:Lowdin}
\end{equation}
Let $|g_n\rangle$ denote the $n$th occupied eigenstate of the flake Hamiltonian. Before orthonormalization, the occupied-state weight of each local $B$ trial is
\begin{equation}
q_i=\sum_{n\in\mathrm{occ}}|\langle g_n|i,B\rangle|^2
=\langle i, B|P|i, B\rangle=S_{ii}.
\label{eq:projectability}
\end{equation}
For the $16\times16$ flake at the representative Chern point
$(E_0/t_2,\varphi)=(2,\pi/2)$, $q_i$ ranges from $0.60$ to $0.75$,
with an all-site mean $0.62$ and a central-site value $0.62$.
These values establish substantial finite occupied-subspace weight for the local $B$ trials used in the finite-flake construction.

\clearpage

\section{Orbital magnetization}
\label{sec:om-benchmark}

For a localized orbital $|w_i\rangle$ with center $\bar{\bm r}_i=\langle w_i|\bm r|w_i\rangle$, the finite-sample orbital moment can be resolved as
\begin{align}
 \bm m_i^{\rm LC}
 &=-\frac{e}{2}\langle w_i|
 (\bm r-\bar{\bm r}_i)\times\bm v|w_i\rangle,
 \label{eq:MLC}\\
 \bm m_i^{\rm IC}
 &=-\frac{e}{2}\bar{\bm r}_i\times
 \langle w_i|\bm v|w_i\rangle.
 \label{eq:MIC}
\end{align}
LC and IC separate self-rotation around the localized center from circulation of that center.
Summing Eqs.~\eqref{eq:MLC} and \eqref{eq:MIC} over every occupied orbital and dividing by the full sample area gives the total flake magnetization.

For each phase and each displayed $E_0/t_2$, the calculation diagonalizes the $16\times16$ open-flake Hamiltonian, sorts the eigenvalues, and occupies the lowest $N_{\rm flake}^2=256$ of the $2N_{\rm flake}^2=512$ one-particle states, where $N_{\rm flake}=16$ is the number of primitive cells along each flake direction.
With $A_0=\sqrt3 a^2/2$ and $A=N_{\rm flake}^2A_0$, the orbital magnetization is
\begin{align}
 M^{\rm tot}&=\frac{1}{A}\sum_i
  \left(m_i^{\rm LC}+m_i^{\rm IC}\right)
  =M_{\rm LC}^{\rm all}+M_{\rm IC}^{\rm direct},
 \label{eq:Mfinite}\\
 M_{\rm LC}&=\frac{m_c^{\rm LC}}{A_0},\qquad
 M_{\rm IC}^{\rm res}=M^{\rm tot}-M_{\rm LC}.
 \label{eq:figurefinite}
\end{align}
Here $c$ denotes a L\"owdin orbital near the flake center, and $\mathrm{res}$ denotes the residual after subtracting $M_{\rm LC}$ from $M^{\rm tot}$. Because
$M_{\rm LC}^{\rm all}$ includes boundary orbitals, Figs.~1(c) and 2(a) use
$M_{\rm LC}$ as the finite bulk-LC estimator. Figure~2(c) uses the
operational residual $M_{\rm IC}^{\rm res}$, and Fig.~2(d) uses
$M^{\rm tot}$~\cite{Thonhauser2005}.
At finite $N_{\rm flake}$ the
operational residual is not assumed to equal the direct all-orbital quantity
$M_{\rm IC}^{\rm direct}$.

The conversion to the physical units used on the finite-flake axes follows from
\begin{equation}
 \frac{m_e}{\hbar^2}(1\ {\rm eV})(1\ \text{\AA})^2
 =0.13,
 \qquad
 \frac{m_e t_1a^2}{\hbar^2}=3.28
 \quad(t_1=1\ {\rm eV},\ a=5\ \text{\AA}).
 \label{eq:moment-conversion}
\end{equation}
When $b_i^z$ is reported in \AA$^{-2}$,
\begin{equation}
 B_{{\rm eq},i}^z[\mathrm T]
 =\frac{\hbar}{e}\,b_i^z[\text{\AA}^{-2}]\,10^{20}
 \simeq 6.58\times10^4\,b_i^z[\text{\AA}^{-2}],
 \label{eq:field-conversion}
\end{equation}
because $1\ \text{\AA}^{-2}=10^{20}\ \mathrm m^{-2}$.  Thus $b^z$ has
inverse-area units, whereas the plotted $B_{\rm eq}^z$ is in tesla.
This equivalent-field label is a dimensional representation of the
molecular Berry curvature.  It is neither an applied laboratory field nor a
calibration of the Haldane hopping phase $\varphi$ in tesla.  The phase
controls internal time-reversal breaking in the electronic reference model;
its relation to an external field is not specified here.

The periodic formulas are used only to validate the finite implementation in the trivial regime. For an occupied band, the Gaussian-unit formulas are~\cite{Thonhauser2005,Ceresoli2006}
\begin{align}
 M_{\rm LC}^{\rm per}
 &=\frac{e}{2\hbar c}\operatorname{Im}\int_{\BZ}
 \frac{\dd^2k}{(2\pi)^2}
 \langle\partial_{\bm k}u_{\bm k}|\times H_{\bm k}
 |\partial_{\bm k}u_{\bm k}\rangle,
 \label{eq:periodic-MLC}\\
 M_{\rm IC}^{\rm per}
 &=-\frac{e}{2\hbar c}\int_{\BZ}\frac{\dd^2k}{(2\pi)^2}
 \epsilon_{\bm k}\,\Omega_{\bm k}.
 \label{eq:periodic-MIC}
\end{align}
As a separate zero-Chern literature benchmark, I use $t_1=1$, $t_2=1/3$,
$E_0=2$, and $\varphi=\pi/2$, rather than the $t_2=t_1/4$ parameters used
for the manuscript figures.  In this benchmark, $M_{\rm LC}=5.67\times10^{-2}$
for $N_{\rm flake}=30$, the quadratic $1/N_{\rm flake}$ extrapolation of $M_{\rm LC}^{\rm all}$ is
$5.67\times10^{-2}$, and the $121\times121$ periodic value is
$M_{\rm LC}^{\rm per}=5.67\times10^{-2}$.  These three values use the same
prefactor-one convention, in which the common physical prefactor multiplying
the LC expressions is set to unity; the comparison tests the finite
normalization rather than the physical axis units of Figs.~1 and 2.

\clearpage

\section{Spatial distribution of molecular Berry curvature}
\label{sec:spatial}

For site $i$, the local molecular Berry curvature that generates the orbital-magnetization--phonon (OMP) Raman coupling is
\begin{equation}
 b_i^z\equiv \mathcal F_{i x,i y}.
 \label{eq:local-b}
\end{equation}
The $16\times16$ profile at $E_0/t_2=8$ and
$\varphi=0.3\pi$ contains 512 sites.  The displayed bulk region is the central
$8\times8$-cell core (128 sites), and the complementary boundary region
contains 384 sites.  Figure~1(d) plots the bulk and boundary
distributions.  The all-site quantity is not drawn as a third histogram; only
its overall mean is retained as a dashed reference line.
Table~\ref{tab:hist-stats} records all three numerical
summaries.

\begin{table}[H]
 \centering
 \caption{Statistics of the displayed equivalent-field coefficient across
 one clean open flake. The boundary region is the complement of the central  $8\times8$-cell core displayed as "Bulk."  Std. is the population standard deviation of all sites in
 the stated region.}
 \label{tab:hist-stats}
 \begin{tabular}{lrrrr}
 \toprule
 Region & Sites & Mean (T) & Std. (T) & Range (T)\\
 \midrule
 All sites & 512 & 9.52 & 2.92 & $0$--13.48\\
 Bulk & 128 & 10.40 & 1.45 & 8.95--11.86\\
 Boundary & 384 & 9.23 & 3.21 & $0$--13.48\\
 \bottomrule
 \end{tabular}
\end{table}

The distinct bulk and boundary histograms, together with the displaced
overall mean, establish spatial nonuniformity.

Here the time-reversal-breaking hopping phase generates electronic orbital circulation. The onsite-disorder test in Sec.~\ref{sec:random-impurity-response} and Fig.~\ref{fig:random-impurity-spread} shows how the local circulation changes when the electronic environment is perturbed. This supports a model with spatially varying local electronic response, while the associated variation of the OMP coupling coefficient is specified phenomenologically in the transport calculation.

The scattering model consists of OMP scattering centers at random positions $\bm R_a$, each carrying a local $b_a^z$.
The configurational average is taken over the random positions; it removes cross-center phase terms and produces momentum relaxation.
Conversely, a coefficient repeated perfectly in every primitive cell
would be periodic and would belong in the phonon Hamiltonian, which is different from the dilute model of OMP scattering centers.

The electronic calculation supplies a dimensionless profile of the molecular Berry curvature. The bulk-mean equivalent fields, proportional to the sublattice means of $b_i^z$, are
$x_A=8.95$ T and $x_B=11.85$ T.  Their mean and normalized
contrasts are
\begin{equation}
 \bar x=\frac{x_A+x_B}{2}=10.40~\mathrm{T},
 \qquad
 s_A=\frac{x_A}{\bar x}=0.86,
 \qquad
 s_B=\frac{x_B}{\bar x}=1.14.
 \label{eq:profile-ratio}
\end{equation}

In the transport model, I use these normalized curvature weights as the relative magnitudes and signs of the sublattice transport coupling energies, $\epsilon_\mu^{\mathrm{tr}}=\epsilon_\star s_\mu$.
This choice fixes the sublattice dependence of the effective vertex.
The bare BO energies instead obey $\epsilon_\mu^{\mathrm{BO}}\propto b_\mu^z/M_\mu$, as derived in Sec.~\ref{sec:BO}; their ratio also depends on the ionic masses.

\clearpage

\section{Evolution of local molecular Berry curvature}
\label{sec:realspace}

The real-space evolution is evaluated on a $16\times16$ open flake with
$t_1=1$ eV, $t_2=0.25$ eV, and $g=0.1$ eV\,\AA$^{-1}$.  For visual comparison only, the
curvature is labeled by the equivalent field
$B_{\rm eq}=(\hbar/e)b$.

\begin{figure}[H]
 \centering
\includegraphics[width=0.70\linewidth]{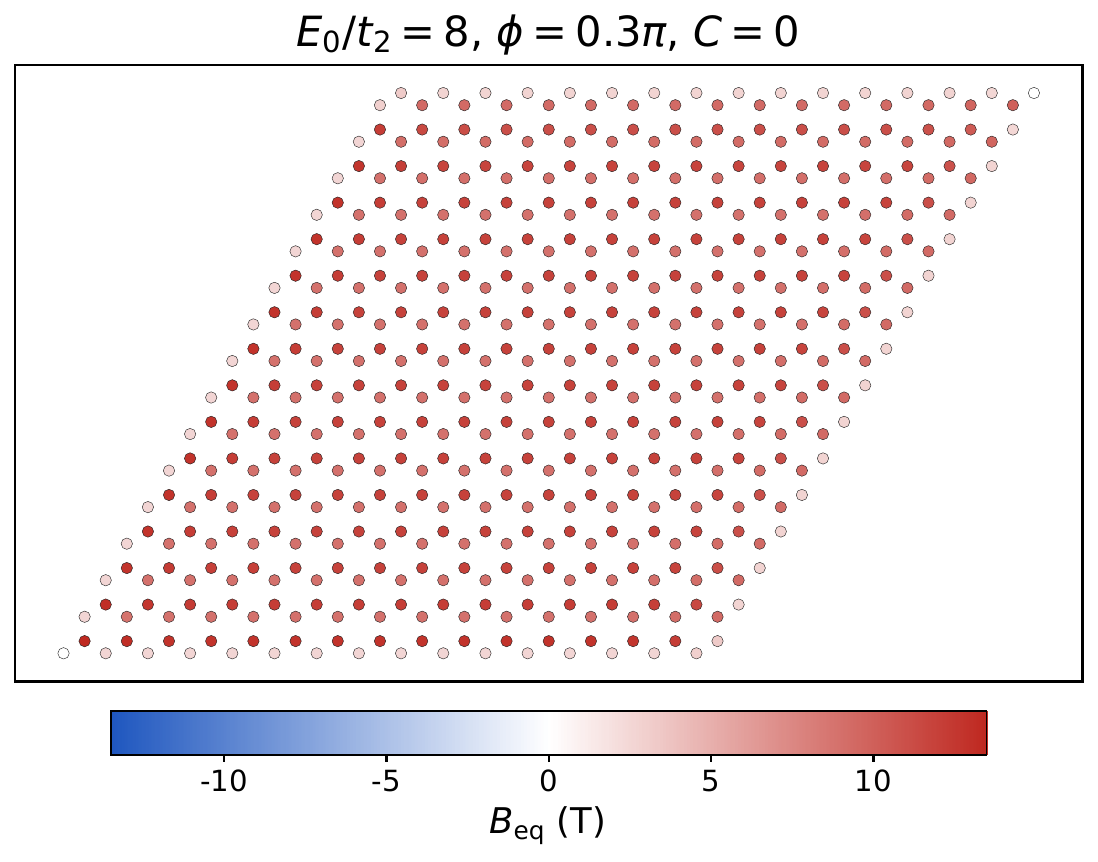}
 \caption{Real-space profile of $B_{\rm eq}$ at $E_0/t_2=8$ and
 $\varphi=0.3\pi$.  This is the spatial source of Fig.~1(d) in the manuscript. Circle color gives the sign and circle size gives the magnitude of the local coefficient.}
 \label{fig:single-map}
\end{figure}

In the displayed trivial regime, $B_{\rm eq}$ evolves smoothly, increasing in
magnitude toward $\varphi=\pi/2,3\pi/2$ and decreasing toward
$\varphi=0,\pi$.  Figure~\ref{fig:single-map} shows a representative
distribution at $E_0/t_2=8$ and $\varphi=0.3\pi$.

\begin{figure}[H]
 \centering
\includegraphics[width=0.98\linewidth]{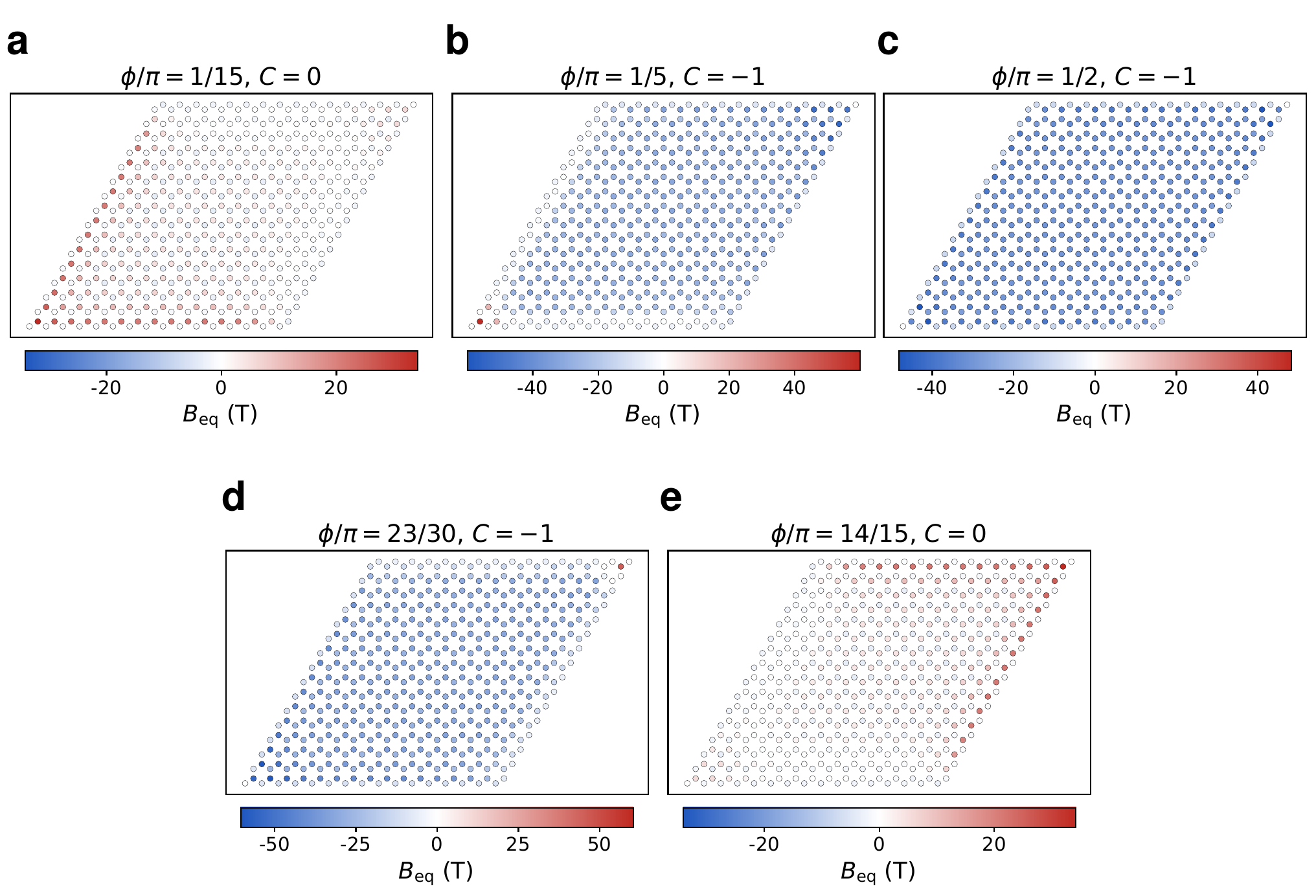}
 \caption{$E_0/t_2=2$ sweep by changing $\varphi/\pi=1/15$, $1/5$, $1/2$, $23/30$, and $14/15$.
 The corresponding Chern numbers are $C=0,-1,-1,-1,0$.
 The infinite-system phase boundaries occur at $\varphi/\pi=0.13$ and $0.87$.
 The interior pattern changes most strongly on crossing the phase boundary and varies more weakly within the deep-Chern regime, directly connecting the spatial evolution to the plateau-like curve included in the main text.
 }
 \label{fig:fixed-sweep}
\end{figure}

\begin{figure}[H]
 \centering
 \includegraphics[width=0.98\linewidth]{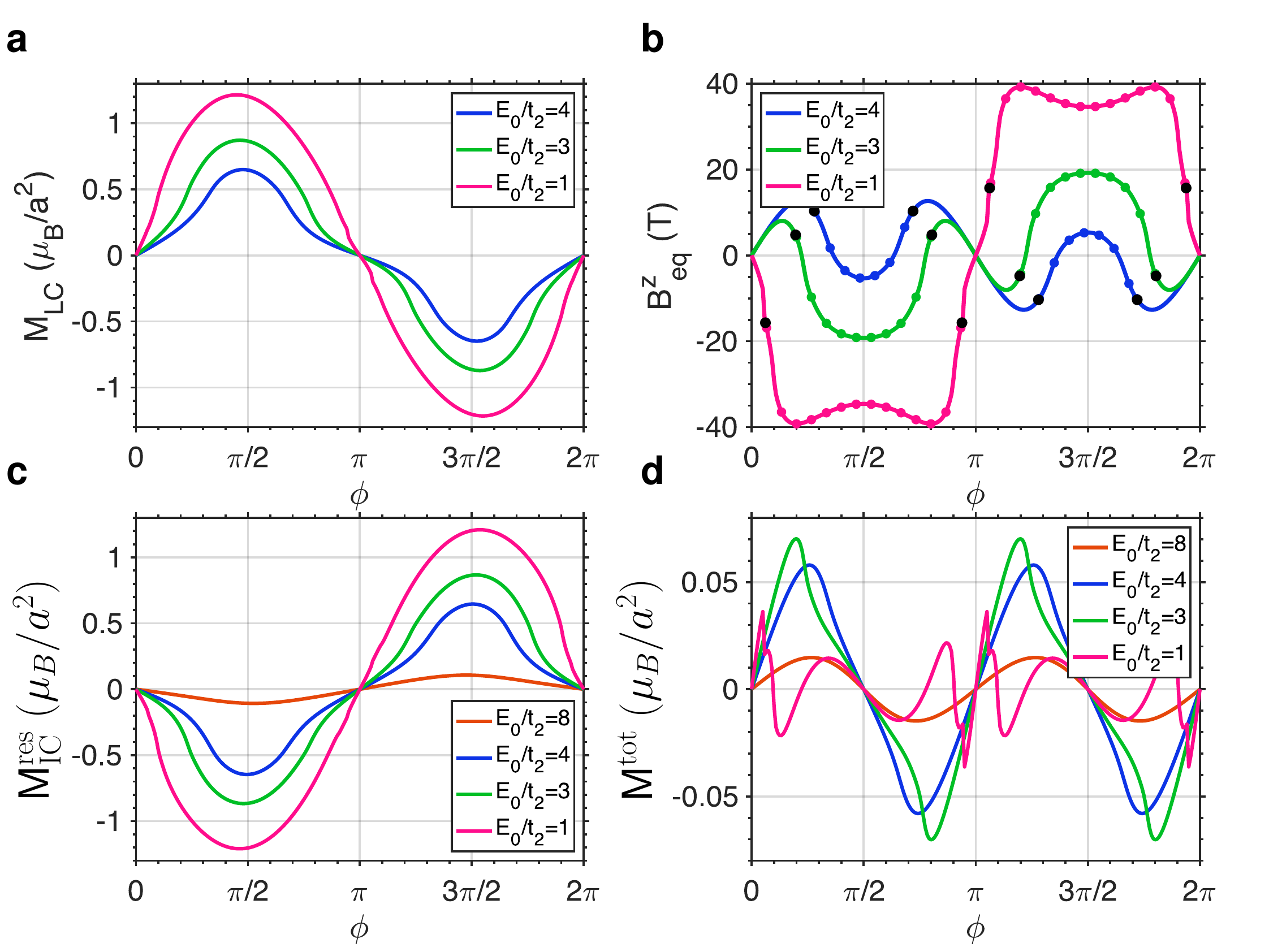}
 \caption{Additional $E_0/t_2$ dependence corresponding to Fig.~2(a)--(d).
 (a),(b) Finite central-orbital $M_{\rm LC}$ and spatially averaged finite-flake
 $B_{\rm eq}^z$ for $E_0/t_2=1,3,4$.  (c),(d) The finite residual
 $M_{\rm IC}^{\rm res}$ and finite total $M^{\rm tot}$ for
 $E_0/t_2=1,3,4,8$.  The definitions, $16\times16$ open flake,
 exact half-filling prescription, and
 241-point phase grid are identical to Fig.~2(a)--(d).
 In (b), the black dots indicate the phase boundary, and the colored dots indicate the Chern regime.}
 \label{fig:additional-ratios}
\end{figure}

Across the topological boundary, the dominant redistribution first appears in
the interior, followed by changes in the boundary profile.  The fixed-ratio
sweep in Fig.~\ref{fig:fixed-sweep} compares the near-boundary and deep-Chern
regimes.  This redistribution coincides with, but is not by itself proof of
the microscopic origin of, the sign reversal and plateau-like trend of
$B_{\rm eq}$ in Fig.~2 of the manuscript.  Similar qualitative features occur
for $E_0/t_2=1,3,4$ as well as for the displayed ratio 2; see
Fig.~\ref{fig:additional-ratios}.

\clearpage

\subsection{Random-impurity spatial response}
\label{sec:random-impurity-response}

\begin{figure}[H]
 \centering
 \includegraphics[width=0.98\linewidth]{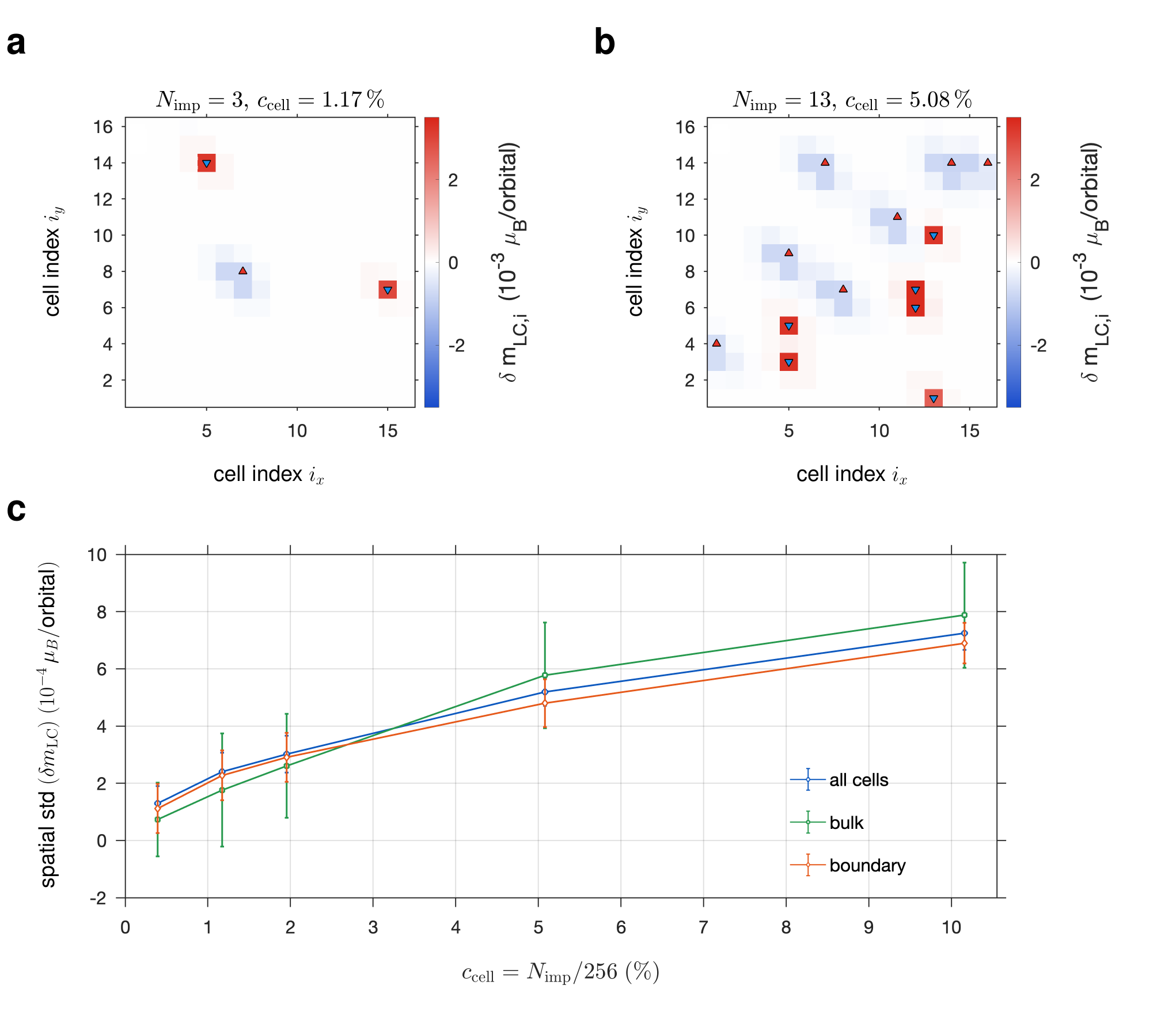}
 \caption{Spatial change in local orbital circulation induced by random onsite impurities.
 (a),(b) Representative $\delta m_i^{\rm LC}$ maps for $N_{\rm imp}=3$ ($c_{\rm cell}=1.17\%$) and $13$ ($5.08\%$).
 Red upward and  blue downward triangles mark onsite perturbations on the $A$ and $B$ sublattices, respectively.
 (c) Configuration-averaged population standard deviation over all cells, the central $8\times8$-cell bulk, and the complementary boundary.
 Error bars are sample standard deviations among the 48 configurations, not standard errors.
 Here $c_{\rm cell}=N_{\rm imp}/256$.
 The plotted quantity is $\delta m_i^{\rm LC}$ per centered B-trial L\"owdin orbital.}
 \label{fig:random-impurity-spread}
\end{figure}

At $(E_0/t_2,\varphi)=(8,0.3\pi)$, I selected
$N_{\rm imp}=1,3,5,13,26$ distinct cells of the $16\times16$ open flake and
applied a real onsite shift $V=0.10$ eV to one randomly chosen $A$ or $B$ site
in each selected cell.  For 48 fixed-count configurations at every
$N_{\rm imp}$, the occupied subspace and the centered B-trial L\"owdin frame
were reconstructed and the local response
\begin{equation}
 \delta m_i^{\rm LC}
 =m_i^{\rm LC}[H_{\mathcal C}]-m_i^{\rm LC}[H_0]
 \label{eq:random-impurity-lc-response}
\end{equation}
was evaluated.  Figure~\ref{fig:random-impurity-spread} shows that this
response is spatially distributed and that its width changes with the number of perturbed cells. In the representative configuration in Fig.~\ref{fig:random-impurity-spread}(a), the maximum absolute relative change,
$\max_i|\delta m_i^{\rm LC}/m_i^{\rm LC}|$, is $4.32\%$.

This calculation tests the response of local orbital circulation to onsite disorder. It does not compute $\delta b_i^z$ or determine the OMP coupling-coefficient distribution used in transport. In particular, the single-site onsite perturbations here differ from the coherent A+B OMP scattering center specified in Sec.~\ref{sec:vertex}. The disorder calculation motivates spatial modulation, while the transport center and its coefficient moments remain model inputs.

\section{Born--Oppenheimer Raman term and linewidth calibration}
\label{sec:BO}

For nuclear coordinates $\bm u$ and an adiabatic many-electron state
$|\Psi(\bm u)\rangle$, define
\begin{equation}
 \mathcal A_i^\alpha=
 \ii\langle\Psi|\partial_{u_i^\alpha}\Psi\rangle,
 \qquad
 \mathcal F_{i\alpha,j\beta}
 =\partial_{u_i^\alpha}\mathcal A_j^\beta
 -\partial_{u_j^\beta}\mathcal A_i^\alpha.
 \label{eq:BO-connection}
\end{equation}
For the independent-electron Haldane ground state, the adiabatic many-electron
state is the Slater determinant
\begin{equation}
 |\Psi(\bm u)\rangle=\frac{1}{\sqrt{N_{\rm el}!}}
 \det[|\phi_1(\bm u)\rangle,\ldots,|\phi_{N_{\rm el}}(\bm u)\rangle].
 \label{eq:Slater}
\end{equation}
Here $N_{\rm el}$ is the number of occupied electronic orbitals. Orthonormality reduces its connection to the occupied-orbital sum
$\mathcal A_i^\alpha=\ii\sum_{n\in\mathrm{occ}}
\langle\phi_n|\partial_{u_i^\alpha}\phi_n\rangle$ and its curvature to the
occupied-projector expressions in Eqs.~\eqref{eq:Fprojector} and
\eqref{eq:Fspectral}. This is the many-body reduction used in the finite
L\"owdin calculation.
The nuclear kinetic energy is
\begin{equation}
 H_{\rm BO}=\sum_i
 \frac{[\bm P_i-\hbar\bm{\mathcal A}_i(\bm u)]^2}{2M_i}
 +U(\bm u).
 \label{eq:HBO}
\end{equation}
For a local antisymmetric block
$\mathcal F_{ix,iy}=f_i=-\mathcal F_{iy,ix}$, the symmetric gauge
$\bm{\mathcal A}_i=(-f_i u_{iy}/2,f_i u_{ix}/2)$ gives, to first order in
$f_i$,
\begin{equation}
 H_R^{\rm BO}=-\sum_i\frac{\hbar f_i}{2M_i}L_i^z
 =-\sum_i\epsilon_i^{\rm BO}\frac{L_i^z}{\hbar},
 \qquad
 \epsilon_i^{\rm BO}=\frac{\hbar^2f_i}{2M_i}.
 \label{eq:Raman}
\end{equation}

Equation~\eqref{eq:Raman} defines the bare BO energy from the microscopic curvature and ionic mass.
For transport, I retain the operator form
$H_R^{\rm tr}=-\sum_{a,\mu}\epsilon_{a\mu}^{\rm tr}L_{a\mu}^z/\hbar$,
where $a$ labels a scattering center and $\mu=A,B$ its two sublattices. The transport coupling $\epsilon_{a\mu}^{\rm tr}=\epsilon_\star s_\mu$ uses the normalized curvature weights of Eq.~\eqref{eq:profile-ratio}.
Thus the transport and bare BO sublattice ratios differ:
\[
 \frac{\epsilon_A^{\rm tr}}{\epsilon_B^{\rm tr}}=\frac{s_A}{s_B},
 \qquad
 \frac{\epsilon_A^{\rm BO}}{\epsilon_B^{\rm BO}}
 =\frac{\bar b_A^z/M_A}{\bar b_B^z/M_B},
\]

where $\bar b_\mu^z$ is the bulk mean on sublattice $\mu$.
The experimental scale guiding the linewidth calibration comes from acoustic-phonon measurements.
For a $9.8$-meV TA-like mode in Ge, neutron spin-echo measurements infer an intrinsic room-temperature linewidth of $0.110(20)$ meV~\cite{Kulda2001}.
In GaN, attenuation measurements give intrinsic room-temperature mean free paths of $5.3$ and $3.5~\mu$m for LA phonons at $1.06$ and $1.43$ THz, respectively~\cite{Chou2019}.

Guided by these measurements, I use a representative impurity-induced phonon full width at half maximum (FWHM) of $\Gamma_*=0.0119$ meV at $c_{\rm imp}=N_c/N_{\rm cell}=0.01$, where $N_c$ is the number of scattering centers in $N_{\rm cell}$ primitive cells. I attribute this target linewidth to OMP scattering in the model. Applying the same procedure to a particular sample requires separating the OMP contribution from other sources of broadening.

On the $50\times50$ momentum grid, the two acoustic branches give 5,000 modes. I use all 38 modes with phonon energies between 8 and 12 meV to characterize the linewidth near 10 meV. Each mode has its own FWHM. I choose a single coupling magnitude $|\epsilon_\star|$ so that the median of the 38 widths (the mean of the 19th and 20th in ascending order) equals $\Gamma_*$.

For scattering from an initial acoustic mode $i$ to a final mode $f$, the single-center matrix element is $U_{fi}=\epsilon_\star\gamma_{fi}$ (Sec.~\ref{sec:vertex}). The dimensionless factor $\gamma_{fi}$ includes the phonon polarizations, energy ratios, and sublattice weights $s_A=0.86$ and $s_B=1.14$. For $N_k$ momentum points, Fermi's golden rule gives the linewidth kernel
\begin{equation}
 \mathcal K_i=\frac{1}{N_k}\sum_{f\ne i}
 |\gamma_{fi}|^2\,\delta(E_f-E_i).
 \label{eq:linewidth-kernel-new}
\end{equation}
The delta function enforces elastic scattering between modes with energies $E_i$ and $E_f$. Summing the concentration-normalized second-order transition rates $\hat W^{(2)}_{fi}$ of Sec.~\ref{sec:FGR} gives the mode linewidth $\Gamma_i^{\rm imp}$~\cite{Sinitsyn2007}. Matching its unweighted median over the selected modes $\mathcal W$ to $\Gamma_*$ gives $|\epsilon_\star|$:
\begin{align}
 \Gamma_i^{\rm imp}
 &=\hbar c_{\rm imp}\sum_f\hat W^{(2)}_{fi}
 =2\pi c_{\rm imp}\epsilon_\star^2\mathcal K_i,
 \label{eq:linewidth-rate-new}\\
 |\epsilon_\star|
 &=\left[
 \frac{\Gamma_*}
 {2\pi c_{\rm imp}\operatorname{median}_{i\in\mathcal W}\mathcal K_i}
 \right]^{1/2},\qquad
 \mathcal W=\{i:8\le E_i/\mathrm{meV}\le12\}.
 \label{eq:linewidth-fit-new}
\end{align}
Here $\Gamma_i^{\rm imp}=\hbar/\tau_{{\rm pop},i}=2\hbar/\tau_{{\rm amp},i}$, where the lifetimes describe population and amplitude decay, respectively~\cite{Cusco2020}. For the 38 modes,
\begin{equation}
 \operatorname{median}_{i\in\mathcal W}\mathcal K_i
 =1.89\times10^{-3}\ \mathrm{meV}^{-1}.
 \label{eq:linewidth-kernel-value-new}
\end{equation}
This gives $|\epsilon_\star|=10.02$ meV. The electronic orbital state sets the time-reversal-odd sign. For the positive sign used in Fig.~3 of the main text, the rounded coupling and corresponding sublattice values are
\begin{equation}
 \epsilon_\star=10\ \mathrm{meV},\qquad
 \epsilon_A^{\rm tr}=8.61\ \mathrm{meV},\qquad
 \epsilon_B^{\rm tr}=11.39\ \mathrm{meV}.
 \label{eq:direct-eps}
\end{equation}

\section{Phonon model and acoustic-branch restriction}
\label{sec:phonon}

The two-dimensional honeycomb lattice has two in-plane displacement
coordinates on each of two sublattices.  For a NN bond direction
$\hat{\bm d}$, the force tensor is
\begin{equation}
 \Phi(\hat{\bm d})=K_\perp\mathbb I_2+
 (K_\parallel-K_\perp)
 \hat{\bm d}\hat{\bm d}^{\mathsf T}.
 \label{eq:force}
\end{equation}
With the three NN directions
$\hat{\bm d}_1=(\sqrt3/2,1/2)$,
$\hat{\bm d}_2=(-\sqrt3/2,1/2)$, and
$\hat{\bm d}_3=(0,-1)$, the mass-weighted dynamical matrix is
\begin{equation}
 D(\bm k)=
 \begin{pmatrix}
 \Phi_0/M_A & -\Phi_{AB}(\bm k)/\sqrt{M_AM_B}\\
 -\Phi_{AB}^\dagger(\bm k)/\sqrt{M_AM_B} & \Phi_0/M_B
 \end{pmatrix},
 \label{eq:dynamical}
\end{equation}
where $\Phi_0=\sum_{j=1}^3\Phi(\hat{\bm d}_j)$ and
\begin{equation}
 \Phi_{AB}(\bm k)=\Phi(\hat{\bm d}_3)
 +\Phi(\hat{\bm d}_1)e^{-\ii k_1}
 +\Phi(\hat{\bm d}_2)e^{-\ii k_2}.
 \label{eq:offblock}
\end{equation}
Here $k_1=\bm k\cdot\bm a_2$ and $k_2=\bm k\cdot(\bm a_2-\bm a_1)$ are dimensionless, with $\bm k$ the wave vector and $\bm a_1,\bm a_2$ defined in Eq.~\eqref{eq:NNN-directions}.
The diagonal blocks in Eq.~\eqref{eq:dynamical} use the canonical unequal-mass
weighting $\Phi_0/M_A$ and $\Phi_0/M_B$.  I use
$K_\parallel=21.998$ eV\,\AA$^{-2}$,
$K_\perp=5.010$ eV\,\AA$^{-2}$, $M_A=10.81$ u, and
$M_B=14.00$ u~\cite{Michel2009}.  After converting the force constants and
masses to a common unit system, an eigenvalue $\lambda_\nu(\bm k)$ of
$D(\bm k)$ gives $\omega_\nu(\bm k)=\sqrt{\lambda_\nu(\bm k)}$ and
$E_\nu(\bm k)=\hbar\omega_\nu(\bm k)$. Numerically,
$E_\nu[\mathrm{meV}]=64.65\sqrt{\lambda_\nu[\mathrm{eV}\,\text{\AA}^{-2}\,\mathrm{u}^{-1}]}$.
The diagonalization gives four nonnegative-frequency branches.

The thermal-transport calculation applies an acoustic-only low-energy
projection: the initial, final, and intermediate virtual states are all
restricted to the two acoustic branches.  Low temperature suppresses real
optical-phonon populations, but does not by itself eliminate virtual optical
intermediate states; neglecting those intermediates is an additional model
approximation.  The optical branches are shown in the dispersion panel for
orientation but do not enter the collision kernel.  Here $N$ is the number of momentum points along each grid direction. On an $N\times N$ midpoint grid, $N_k=N^2$.  Velocities are obtained from centered differences
with dimensionless step $10^{-4}$ in $k_1$ and $k_2$, and converted using

\begin{equation}
 v_x=\frac{a}{2\hbar}(\partial_{k_1}E-\partial_{k_2}E),
 \qquad
 v_y=\frac{\sqrt3a}{2\hbar}(\partial_{k_1}E+\partial_{k_2}E).
 \label{eq:velocity}
\end{equation}

\section{One-phonon scattering}
\label{sec:vertex}

Let $E_i>0$ be an acoustic-phonon energy and
$z_{\mu i}$ its two-component polarization on sublattice $\mu$.  With the
basis phases $\chi_A=1$ and
$\chi_B=\exp[\ii(k_1+k_2)/3]$, set
$\widetilde z_{\mu i}=\chi_\mu z_{\mu i}$. Quantizing the local Raman
operator $H_R^{\rm tr}$ defined in Sec.~\ref{sec:BO} gives the
number-conserving one-phonon matrix element
\begin{equation}
 U_{fi}=-\frac12\sum_{\mu=A,B}\epsilon_\mu^{\rm tr}
 \left(\sqrt{\frac{E_f}{E_i}}+
       \sqrt{\frac{E_i}{E_f}}\right)
 \widetilde z_{\mu f}^\dagger\sigma_y\widetilde z_{\mu i},
\end{equation}
where $\sigma_y$ is the Pauli matrix acting in the Cartesian $(x,y)$
displacement space. Every term in $U_{fi}$ has units of meV.

The matrix $\sigma_y$ represents phonon angular momentum; time reversal changes $\varphi\mapsto-\varphi$, $b_i^z\mapsto-b_i^z$, and $L_z\mapsto-L_z$, so the odd OMP coefficient comes from the electronic state rather than from an independent degree of freedom.
For a gapped spinless time-reversal-invariant reference with real hopping and displacement derivatives, the spectral curvature vanishes [Eqs.~\eqref{eq:Fspectral} and \eqref{eq:Raman}].

Let $U_{fi}^{(a)}$ denote the single-center internal vertex appearing in each
summand of Eq.~\eqref{eq:random-vertex}, evaluated with the local coefficients
at a center $a$ located at primitive-cell position $\bm R_a$.  For a random
configuration $\mathcal C$, the configuration vertex is the position-phase
sum
\begin{equation}
\mathcal U_{fi}[\mathcal C]=\sum_{a\in\mathcal C}
 e^{\ii(\bm k_i-\bm k_f)\cdot\bm R_a}
 U_{fi}^{(a)}.
 \label{eq:random-vertex}
\end{equation}
For Bloch states normalized in a box of $N_k$ primitive cells, the physical
random-configuration matrix element is
\begin{equation}
 V_{fi}[\mathcal C]=\frac{1}{N_k}\mathcal U_{fi}[\mathcal C].
 \label{eq:box-vertex}
\end{equation}
For nonzero momentum transfer,
\begin{align}
 \big\langle |\mathcal U_{fi}|^2\big\rangle_{\mathcal C}
 &\longrightarrow N_c
 \big\langle |U_{fi}^{(a)}|^2\big\rangle,\\
 \big\langle \mathcal U_{fi}^*\mathcal U_{fv}\mathcal U_{vi}\big\rangle_{\mathcal C}
 &\longrightarrow N_c
 \big\langle (U_{fi}^{(a)})^*U_{fv}^{(a)}U_{vi}^{(a)}
 \big\rangle.
 \label{eq:center-contractions}
\end{align}
Here, $N_c=c_{\rm imp}N_k$. The mean-power approximation applies to
moments of the local coupling coefficients at one center:
\[
 \left\langle\prod_{j=1}^{n}\epsilon_{a\mu_j}^{\rm tr}\right\rangle_{\epsilon}
 \simeq\prod_{j=1}^{n}\epsilon_{\mu_j}^{\rm tr},\qquad
 n=2,3,\quad \mu_j\in\{A,B\}.
\]
The average $\langle\cdot\rangle_\epsilon$ concerns local-coefficient variation.
The positional average in Eq.~\eqref{eq:center-contractions} is performed on
the quadratic and cubic products before making this approximation.

The average over independent random center positions restores translation invariance, leaving the finite-momentum-exchanging matrix elements.
For independent dilute OMP scattering centers at a scattering-center concentration $c_{\mathrm{imp}}$ per primitive cell,
\begin{equation}
 W_{fi}^{(n)}(c_{\mathrm{imp}})
 =c_{\mathrm{imp}}\hat W_{fi}^{(n)}+O(c_{\mathrm{imp}}^2),
 \qquad n=2,3.
 \label{eq:dilute-expansion}
\end{equation}
The common linear factor is the standard scaling of the quadratic rate and the conventional third-order skew-scattering rate for independent scattering centers~\cite{Sinitsyn2007}.
I report $\hat W^{(n)}$ obtained after dividing out this common factor.

\section{Fermi golden rule and third-order scattering rate}
\label{sec:FGR}

Within the Born approximation, the standard $T$ matrix is
\begin{equation}
 T_{fi}=V_{fi}+
 \sum_v\frac{V_{fv}V_{vi}}
 {E_i-E_v+\ii\eta}+O(V^3).
 \label{eq:Tborn}
\end{equation}
Using $N_c=c_{\mathrm{imp}}N_k$, Eqs.~\eqref{eq:box-vertex} and
\eqref{eq:center-contractions}, and then dividing out $c_{\mathrm{imp}}$ gives
the concentration-normalized rate matrices stored on a uniform $N_k$-point mesh,
\begin{align}
 \hat W_{fi}^{(2)}
 &=\frac{2\pi}{\hbar}q_f |U_{fi}|^2
 \delta(E_f-E_i),
 \label{eq:W2}\\
 \hat W_{fi}^{(3)}
 &=\frac{4\pi}{\hbar}q_f\,
 \mathrm{Re}\!\left[
 U_{fi}^*\sum_vq_v
 \frac{U_{fv}U_{vi}}{E_i-E_v+\ii\eta}\right]
 \delta(E_f-E_i),
 \qquad q_f=q_v=\frac1{N_k}.
 \label{eq:W3}
\end{align}
The diagonal entries are set to $\hat W_{ii}^{(2)}=\hat W_{ii}^{(3)}=0$;
only transitions to a different state contribute to the transport out-rate.
Reversing the time-reversal-odd OMP coupling coefficients,
$\epsilon_\mu^{\rm tr}\mapsto-\epsilon_\mu^{\rm tr}$, gives $U\mapsto-U$.
Thus, $\hat W^{(2)}$ is even and $\hat W^{(3)}$ is odd under this reversal.
The antisymmetric part of the skew-scattering operator is
\begin{equation}
 \hat W^{(3a)}=\frac12\left(\hat W^{(3)}-[\hat W^{(3)}]^{\mathsf T}\right).
 \label{eq:Wa}
\end{equation}
I group phonon energies into bins of width $\Delta=\pi\eta$, where $\eta$ is the numerical energy resolution.
Let $s$ be the bin containing the initial energy $E_i$ and $E_s$ its center.
I approximate $\delta(E_f-E_i)$ by $1/\Delta$ if the final energy $E_f$ lies in $s$ and by zero otherwise.
To reduce sensitivity to the placement of bin boundaries, I average the rates over $n_{\rm shift}$ boundary positions uniformly spaced within one bin width.
The sum over virtual states $v$ of energy $E_v$ in Eq.~\eqref{eq:W3} is evaluated as
\[
 \sum_{v\notin s}q_v U_{fv}U_{vi}
 \frac{E_s-E_v}{(E_s-E_v)^2+\eta^2}
 -\frac{\ii\pi}{\Delta}\sum_{v\in s}q_v U_{fv}U_{vi},
\]
with the principal-value contribution from states in $s$ set to zero.

\section{Linearized Boltzmann equation}
\label{sec:BTE}

Write the nonequilibrium occupation as
$n_i=n_i^0+g_i\partial_xT$.  The driving term is
\begin{equation}
 d_i=v_{ix}\frac{\partial n_B(E_i,T)}{\partial T},
 \qquad
 \frac{\partial n_B}{\partial T}
 =\frac{E_i}{k_BT^2}
 \frac{e^{-E_i/k_BT}}{(1-e^{-E_i/k_BT})^2}.
 \label{eq:drive}
\end{equation}
For a transition matrix $W$, the collision matrix is
\begin{equation}
 C_{fi}=\delta_{fi}\sum_jW_{ji}-W_{fi}.
 \label{eq:collision}
\end{equation}
Because the elastic collision matrix has a number-conserving null mode, the
linear system determines $g$ only up to that homogeneous component.  I select
a unique representative by imposing $\sum_i g_i=0$, corresponding to zero net
nonequilibrium occupation; this constraint fixes the null-mode gauge rather
than supplying the microscopic statement of number conservation.

The transport response is evaluated at leading skew order for the
concentration-normalized collision kernels,
\begin{equation}
 \hat C_2\hat g^{(0)}=-d,
 \qquad
 \hat C_2\hat g^{(1)}=-\hat C_a\hat g^{(0)}.
 \label{eq:perturbative}
\end{equation}
Here $\hat g_i=c_{\mathrm{imp}}g_i$ is the concentration-normalized occupation response, $\hat C_2=C[\hat W^{(2)}]$, and $\hat C_a=C[\hat W^{(3a)}]$. The longitudinal response is obtained from $\hat g^{(0)}$, and the skew transverse response from $\hat g^{(1)}$.
This includes the second-order symmetric and the third-order antisymmetric scattering rates in the Born expansion.
The time-reversal-odd component of the concentration-normalized conductivity is given by
\begin{equation}
 \hat\kappa_{\alpha x}^{\rm odd}
 =\frac{\hat\kappa_{\alpha x}(+\epsilon^{\rm tr})
       -\hat\kappa_{\alpha x}(-\epsilon^{\rm tr})}{2}.
 \label{eq:Raman-odd}
\end{equation}

The concentration-normalized conductivity is
\begin{equation}
 \hat\kappa_{\alpha x}
 \equiv\lim_{c_{\mathrm{imp}}\to0}c_{\mathrm{imp}}\kappa_{\alpha x}
 =-\frac{1}{N_k\Omega_3}
 \sum_i E_i v_{i\alpha}\hat g_i,
 \qquad
 \Omega_3=\Omega_2d,
 \qquad
 \Omega_2=\frac{\sqrt3}{2}a^2.
 \label{eq:kappa}
\end{equation}
Here $d=5$ \AA\ is the effective thickness of the system. Both $\hat\kappa_{xx}$ and $\hat\kappa_{yx}^{\rm odd}$ scale as $1/d$, so the Hall angle $\theta_H^{\rm odd}=\hat\kappa_{yx}^{\rm odd}/\hat\kappa_{xx}$ is independent of this conversion.
Assuming that dilute OMP scattering centers provide the only relaxation channel gives $\kappa_{\alpha x}=\hat\kappa_{\alpha x}/c_{\mathrm{imp}}+O(1)$, so the common concentration factor also cancels from $\theta_H$.

\section{Numerical parameters and convergence}
\label{sec:validation}

The transport calculation uses $N=50$ ($N_k=2500$),
$\eta=3$ meV ($\Delta=9.42$ meV), $n_{\rm shift}=128$ bin origins as
defined in Sec.~\ref{sec:FGR}, the two acoustic branches only, and
$T=5$--120 K in 5-K steps.  At 100 K it gives
\begin{equation}
 \hat\kappa_{xx}=1.55\ \mathrm{W\,m^{-1}K^{-1}},\qquad
 \hat\kappa_{yx}^{\rm odd}=1.25\times10^{-3}
 \ \mathrm{W\,m^{-1}K^{-1}},
 \label{eq:kappa100}
\end{equation}
and
\begin{equation}
 \theta_H^{\rm odd}=8.09\times10^{-4}.
 \label{eq:theta100}
\end{equation}

\begin{figure}[H]
 \centering
 \includegraphics[width=0.96\linewidth]{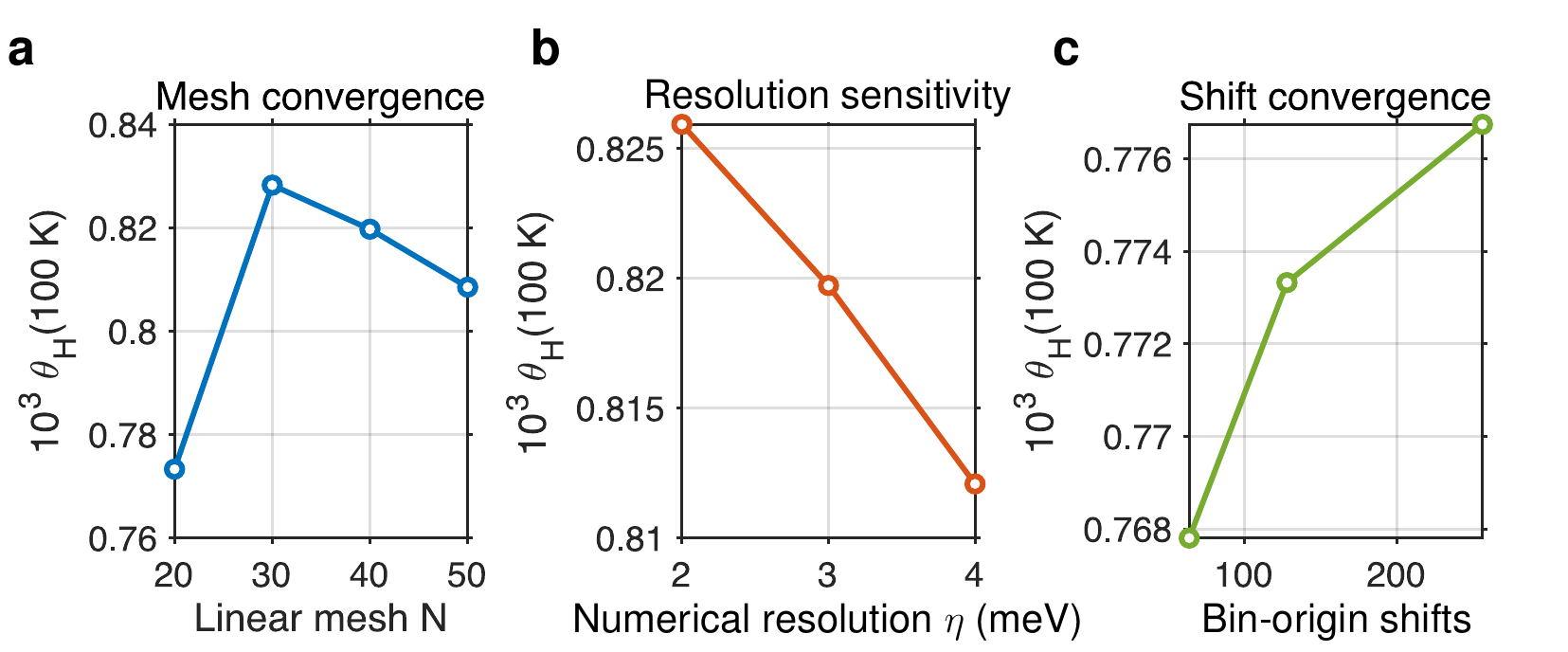}
 \caption{Mesh, numerical energy resolution, and bin-origin-shift checks for the calculated thermal Hall angle.
 (a) At 100 K and $\eta=3$ meV, $\theta_H^{\rm odd}=(7.73,\,8.28,\,8.20,\,8.09)\times10^{-4}$
 for $N=20,30,40,50$, respectively.
 (b) At $N=40$, changing $\eta$ from 2 to 4 meV yields $(8.26,\,8.12)\times10^{-4}$ at the two endpoints.
 (c) At $N=20$ and $\eta=3$ meV,  64, 128, and 256 shifts yield $(7.68,\,7.73,\,7.77)\times10^{-4}$, respectively.}
 \label{fig:convergence}
\end{figure}

Figure~\ref{fig:convergence} shows the mesh, numerical energy resolution, and bin-origin-shift
checks at fixed $\epsilon_\star=10$ meV. The mesh ranges from $20\times20$ to $50\times50$, $\eta$ from 2 to
4 meV, and the number of bin-origin shifts from 64 to 256.  The Hall angle
changes by about $1.37\%$ between the $40\times40$ and $50\times50$ meshes, by
about $1.68\%$ between $\eta=2$ and 4 meV, and by about $0.44\%$ between 128
and 256 shifts.

\begin{table}[H]
 \centering
 \caption{Numerical parameters used for Fig.~3.}
 \label{tab:parameters}
 \begin{tabularx}{\linewidth}{@{}l l X@{}}
 \toprule
 Quantity & Value & Status\\
 \midrule
 $a$ & 5.0 \AA & fixed\\
 $d$ & 5.0 \AA & 3D reporting convention\\
 $K_\parallel$ & 21.998 eV\,\AA$^{-2}$ & fixed\\
 $K_\perp$ & 5.010 eV\,\AA$^{-2}$ & fixed\\
 $M_A,M_B$ & 10.81 u, 14.00 u & fixed\\
 $\epsilon_A^{\rm tr},\epsilon_B^{\rm tr}$ & 8.61, 11.39 meV & \leavevmode calibrated using a target impurity-induced phonon FWHM guided by experimental acoustic-phonon linewidths; see Sec.~\ref{sec:BO}\\
 $c_{\mathrm{imp}}$ & dilute limit & concentration-normalized kernels and conductivities\\
 Acoustic scope & branches 1 and 2 & acoustic-only projection for initial/final/virtual states\\
 Mesh & $50\times50$ & Fig.~3\\
 $\eta$ & 3 meV & Fig.~3\\
 Shift average & 128 origins & Fig.~3\\
 Temperature & 5--120 K & 5-K step\\
 \bottomrule
 \end{tabularx}
\end{table}

\begin{figure}[H]
\centering

\includegraphics[width=0.70\linewidth]{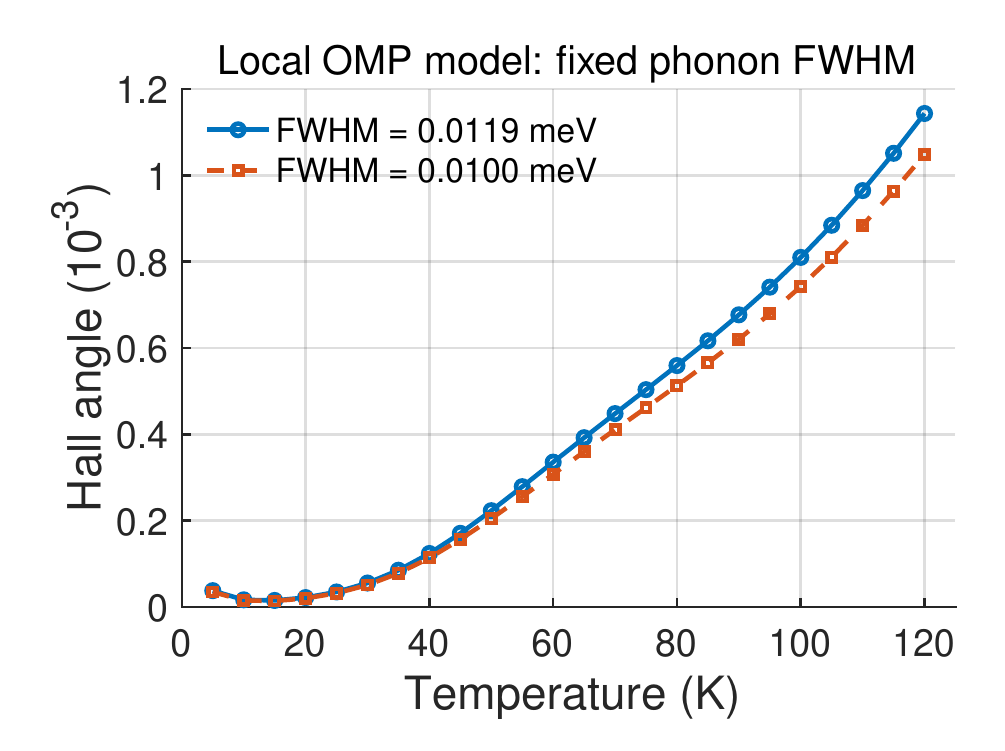}
\caption{Thermal Hall angle for the two target linewidths in Table~\ref{tab:linewidth-direct-new}. The model parameters are fixed as the Bose occupations vary with temperature (5--120 K in 5-K steps).}
\label{fig:linewidth-transport-new}

\end{figure}

I also solve the Boltzmann equation on the full acoustic grid for each of the two target linewidths, using the coupling obtained from Eq.~\eqref{eq:linewidth-fit-new} in Sec.~\ref{sec:BO}. Table~\ref{tab:linewidth-direct-new} gives the resulting conductivities and Hall angles at 100 K, and Fig.~\ref{fig:linewidth-transport-new} shows the Hall angles as functions of temperature.
\begin{table}[H]
\centering

\caption{Direct transport calculations at fixed impurity-induced phonon FWHM.  Values
are at $100$ K, $c_{\rm imp}=0.01$, $d=5$ \AA, $N=50$,
$\eta=3$ meV, and 128 shifted bin origins.
Conductivities are in $\mathrm{W\,m^{-1}K^{-1}}$.
The positive sign uses the Raman convention in Eq.~\eqref{eq:direct-eps}
and $j_y=-\kappa_{yx}\partial_xT$.}
\label{tab:linewidth-direct-new}
\begin{tabular}{rrrrr}
\toprule
$\Gamma_*$ (meV)&$\epsilon_\star$ (meV)&$\kappa_{xx}$&$\kappa_{yx}^{\rm odd}$&$\theta_H^{\rm odd}$\\
\midrule
0.0119&10.02&154.65&$1.25\times10^{-1}$&$8.10\times10^{-4}$\\
0.0100&9.18&184.03&$1.37\times10^{-1}$&$7.43\times10^{-4}$\\
\bottomrule
\end{tabular}

\end{table}
At fixed concentration, the results follow $\kappa_{xx}\propto\epsilon_\star^{-2}$, $\kappa_{yx}^{\rm odd}\propto\epsilon_\star^{-1}$, and $\theta_H^{\rm odd}\propto\epsilon_\star$.

\section*{Supplementary Data 1. Numerical source data}
The archive contains numerical data for the electronic structure, phonon dispersion, linewidth calibration, and thermal transport. The accompanying index identifies the source data for individual figures and tables.